\documentclass[aps,floatfix,onecolumn,a4paper,noshowpacs, nofootinbib,superscriptaddress,11pt]{revtex4}

\usepackage{graphicx,float,tikz}
\usepackage[all]{xy}
\usepackage{physics,bm,amsmath,upgreek,bigints}
\usepackage{amssymb}
\usepackage{color}
\usepackage{epsfig,bm}		
\usepackage{graphicx,epstopdf}
\usepackage{subfigure}
\usepackage{pdfpages}
\usepackage{multirow}
\usepackage{array}
\newcommand{\clt}{\textcolor{black}}

\usepackage{bookmark,textgreek}
\usepackage{hyperref,color,xcolor}
\hypersetup{hidelinks,colorlinks=true,breaklinks=true,urlcolor= blue}
\hypersetup{%
    colorlinks = true,
    linkcolor  = blue,
    citecolor = cyan,
  }

\usepackage{natbib}
\setcitestyle{square,numbers}

\newcommand\orcidroldao{{\href{https://orcid.org/0000-0003-3978-532X}{\orcidicon}}}
\newcommand\orcidpedro{{\href{https://orcid.org/0000-0001-9151-0900}{\orcidicon}}}

\newcommand{\orcidicon}{%
	\begin{tikzpicture}
	\draw[lime, fill=lime] (0,0)
		circle [radius=0.16]
		node[white] {{\fontfamily{qag}\selectfont \tiny ID}};
	\draw[white, fill=white] (-0.0625,0.095)
		circle [radius=0.007];
	\end{tikzpicture}	\hspace{-2mm}
}

\begin{document}

\title{Deformed AdS/QCD, mesonic mass spectra, and DCE: Still a  margin for heavier resonances}
\author{R. da Rocha\orcidroldao\!\!}
\email{roldao.rocha@ufabc.edu.br}
\affiliation{Federal University of ABC, Center of Mathematics, Santo Andr\'e, 09580-210, Brazil}
\author{P. H. O. Silva\orcidpedro\!\!}
\email{silva.pedro@ufabc.edu.br}
\affiliation{Federal University of ABC, Center of Physics, Santo Andr\'e, 09580-210, Brazil}
\begin{abstract}
The mass spectra of light-flavor  mesons are analyzed in a  deformed AdS/QCD soft-wall model, driven by four distinct anomalous 5-dimensional mass corrections of the
scalar field, coupled to Einstein--Hilbert gravity, from the QCD running coupling. Using the differential configurational entropy (DCE) underlying the families of pseudoscalar, axial-vector, scalar, and vector mesons, the mass spectra of heavier meson resonances with radial quantum numbers beyond the ones already in the summary table of  Particle Data Group (PDG) are then estimated.
This protocol merges  AdS/QCD and experimental data in PDG through Regge-like trajectories. Some of the estimated meson resonances may be identified as further candidates omitted from the summary table in PDG. 
\end{abstract}

\maketitle

\section{Introduction}

Based on the concepts of Shannon’s information theory, configurational information measures (CIMs) were formulated as a way to probe both discrete and continuous physical systems, from their description through appropriate probability distributions and wave modes in momentum space \cite{Gleiser:2018kbq,Gleiser:2018jpd}.
Shannon information entropy provides new observables and methods
 to study physical phenomena from both theoretical and experimental points of view. The Shannon information entropy 
is equivalent to information chaoticity carried by messages. In other words, the smaller the information
entropy of a system, the smaller the chaoticity it contains. It also measures the
loss of information and chaotic evolution  in the dynamical evolution of physical systems, mainly in QCD \cite{Ma:2018wtw}. This advantage makes it a suitable tool for studying 
 the evolution of dynamical processes in physical systems.
One of these CIMs, the differential configurational entropy (DCE), evaluates the amount of information needed to encode the physical system under scrutiny, also estimating the very upper limit of lossless compression of data, encrypted into the so-called information dimension \cite{Bernardini:2016hvx}. 

Inspired by the AdS/CFT correspondence \cite{Maldacena:1997re,Witten:1998qj,Gubser:1998bc}, the AdS/QCD correspondence is a fruitful holographic scenario for applying CIMs, in a setup where the QFT is not a conformal field theory. In the ultraviolet (UV) limit and weak-coupling regime, calculations in QCD using perturbative methods work  well to find results compatible with experimental data. However, the same does not hold in the infrared (IR) limit, corresponding to the strongly-coupled regime of QCD, where chiral symmetry breaking and confinement emerge.  AdS/CFT establishes that fields in a weakly-coupled theory of gravity living in the 5-dimensional AdS bulk are dual to operators in a strongly-coupled conformal theory in four dimensions living on the boundary of the AdS bulk  \cite{Aharony:1999ti}. Consequently, results obtained within a weakly coupled theory, and thus more straightforwardly manageable, can be mapped to solutions of the dual theory in non-perturbative regimes, where calculations would be too intricate to be solved with known methods in lattice QCD.

Therefore, employing holographic correspondence for QCD is of great interest. The additional dimension along the AdS bulk is equivalent to an energy scale in four dimensions, where low [high] energies correspond to the IR [UV] limit.
Nevertheless, since QCD is not exactly conformal, it is still possible to accommodate it on the CFT side of the duality through a bottom-up approach, which constructs the dual theory by imposing constraints that make it capable of reproducing phenomenological results \cite{Afonin:2022hbb,dePaula:2009za,Ballon-Bayona:2017sxa,Ballon-Bayona:2021ibm}. One possible bottom-up approach is the soft-wall AdS/QCD model, characterized by the addition of a scalar dilaton field,  coupled to Einstein--Hilbert gravity, whose aim consists of introducing a smooth cutoff in the AdS bulk, emulating confinement while accurately reproducing Regge trajectories \cite{Karch:2006pv,Csaki,Gherghetta:2009ac}. In this setup, the 
AdS bulk metric is deformed 
due to both the chiral and the gluon condensations and the linear confinement and chiral symmetry breaking can be resolved.  
An advantage of this formulation is that the dilaton can be modified to include new properties in the model, such as a deformation that considers the masses of the constituent quarks and consequently generates non-linear trajectories, also  incorporating hybrid and exotic mesons in AdS/QCD  \cite{MartinContreras:2020cyg,Karapetyan:2021ufz}.

Since the first application of the DCE to analyze mesonic states in AdS/QCD in Ref. \cite{Bernardini:2016hvx}, CIMs have proven to be very efficient in obtaining new properties of hadronic resonances and endorsing other well-established features.
The recent use of the CIMs in AdS/QCD has provided new phenomenological insights into some light- and heavy-flavored meson families. This includes quarkonia, heavy-flavor exotic and multiquark mesons, higher-spin tensor mesons, hybrid mesons, baryonic white matter,  odderons, and glueball fields \cite{Bernardini:2018uuy,Braga:2017fsb,Karapetyan:2018oye,Karapetyan:2018yhm,Braga:2018fyc,Braga:2020myi,daRocha:2021imz,MartinContreras:2022lxl,Correa:2015lla,Karapetyan:2023sfo,MartinContreras:2023eft,Kou:2023azd,daRocha:2023nsb,MartinContreras:2023oqs,MartinContreras:2021bis}.
It also encompasses their description at finite temperatures, the quark-gluon plasma in the presence of a magnetic field, the color-glass condensate, and various other nuances of phenomenological QCD  \cite{Colangelo:2018mrt,Ferreira:2020iry,Braga:2020hhs,Karapetyan:2019fst,Karapetyan:2021vyh,Karapetyan:2020yhs,Frederico:2014bva,Braga:2021fey,Braga:2021zyi,Karapetyan:2016fai,Karapetyan:2021crv,Karapetyan:2017edu,Barbosa-Cendejas:2018mng}.
Furthermore, the DCE has been utilized for probing the gravitational collapse and turbulence in AdS space \cite{Barreto:2022ohl}, to scrutinize quantum aspects of compact objects \cite{Casadio:2023pmh,Casadio:2022pla,Koliogiannis:2022uim}, as well as to explore other key aspects of AdS/CFT \cite{Braga:2019jqg,Cruz:2019kwh,Lee:2019tod,Bazeia:2018uyg,Bazeia:2021stz,Braga:2020opg,Lee:2018zmp}. The Higgs boson mass was also determined using the DCE as well as the axion mass in the low energy regime of effective field theory \cite{Alves:2017ljt,Alves:2014ksa}.
In particular, the use of the DCE in the mass spectroscopy of several meson families has already yielded results validated by the Particle Data Group (PDG) \cite{ParticleDataGroup:2024cfk}, and has been shown to produce extrapolated data that coincide with new meson resonance candidates.

One of the possible ways to extend the soft-wall AdS/QCD model consists of the dynamical holographic QCD (DhQCD) model,  known for describing flavor dynamics by coupling it to gluon dynamics \cite{Li:2012ay,Li:2013oda,Huang:2013qvz,Li:2014txa}. The premise is based on the duality between the dilaton field in the bulk and the gluon condensate on the AdS boundary. By complementing the soft-wall model with a gluon dynamics background, DhQCD effectively presents both flavor and gluon dynamics.
The model has been used to study hadron spectra, including glueballs, QCD phase transition, form factors, and transport properties  \cite{dePaula:2008fp,Huang:2013qvz,Li:2013oda,Li:2014txa,Chen:2022goa,Chen:2022pgo}.

In this work, the DCE will be employed within the DhQCD model to compute the mass spectra of resonances with higher values of radial quantum number, in four families of light-flavor mesons. Pseudoscalar $\uppi$ mesons, axial-vector mesons $a_1$, scalar mesons $f_0$, and vector mesons $\rho$ will be analyzed and addressed. These meson families have a considerable number of states and resonances already experimentally detected. Therefore, the results presented in this work suggest the possibility of a margin for the existence of new, heavier resonances yet to be either discovered or identified to compatible further meson states omitted from the summary table in PDG.
Using the DCE, it is possible to obtain DCE-Regge-like trajectories, where the DCE is expressed as a function of the resonance level labeled by the radial quantum number $n$. Similarly, the DCE can be alternatively described as a function of the squared mass of each resonance, generating a second type of DCE-Regge-like trajectories.
Thus, the DCE will be used as an intermediary step to relate the two possible DCE-Regge-like trajectories and to extrapolate the curves to obtain the mass spectra of mesonic resonances for larger values of $n$, beyond those already detected experimentally.

This paper is organized as follows:
Section \ref{section:themodel} introduces the fundamental ingredients of the model, including the basics of the bottom-up soft-wall AdS/QCD and how it is coupled to the gluon dynamics. The model is extended to include four variations of anomalous mass correction, treated as different cases.
Section \ref{section:dce} describes and discusses the DCE protocol, followed by its application to the $\uppi$ pseudoscalar mesons, $a_1$ axial-vector mesons, $f_0$ scalar mesons, and $\rho$ vector mesons, respectively, in subsections \ref{subsection:pi} --  \ref{subsection:rho}. In each subsection, the DCE is calculated and used to construct the two Regge-like trajectories for a given family of mesons, with extrapolations to identify heavier resonances. These results are then compared to further meson states cataloged in PDG in the search for possible pairings.
Section \ref{section:conclusion} discusses and analyzes the relevant results.

\section{The model}
\label{section:themodel}

Following the most common approach to reproduce meson spectroscopy within the soft-wall model, the AdS background bulk metric is written as \cite{Karch:2006pv}
\begin{flalign}
    ds^2 = e^{2A(z)}(\eta_{\mu\nu}dx^\mu dx^\nu + dz^2),
    \label{adsmetric}
\end{flalign}
where $A(z)=\log(L/z)$ stands for the warp factor, $L$ denotes the curvature radius of AdS, and $\eta_{\mu\nu}$ is the Minkowski metric corresponding to the 4-dimensional spacetime representing the boundary of AdS.
The holographic coordinate $z$ is highlighted for convenience, as it can be viewed as an energy scale of the model. Throughout this work, Greek letters indicate the four-dimensional spacetime coordinates ($\mu=0,1,2,3$), while uppercase Latin letters label all five-dimensional coordinates, including the holographic one, e.g., $x^M=(x^\mu,z)$. 
The standard dilaton is a quadratic function of the energy scale, $\phi(z) = \kappa^2 z^2$, having the simplest form that satisfies boundary conditions and accurately reproduces the linear Regge trajectories. The parameter $\kappa$ corresponds to the QCD characteristic energy scale and is usually considered to be around 0.43 GeV \cite{Li:2012ay,Csaki}.

QCD has a SU$(N_f)_L\times {\rm SU}(N_f)_R$ flavor symmetry in the chiral limit, and the model focuses on the lowest-dimensional
operators important to chiral dynamics. These are the left- and right-handed quark currents $\bar{q}_{L}\gamma_{\mu} t^{a} q_{L}$ and $\bar{q}_{R}\gamma_{\mu} t^{a} q_{R}$, where $t^{a}$ are the generators of SU$(N_f)$.  They are associated with the 5-dimensional gauge
fields $A^{a}_{L}$ and $A^{a}_{R}$ using the AdS/CFT dictionary. 
In holographic QCD, the bilinear quark operator $\bar{q}q$, responsible for chiral symmetry breaking in QCD, corresponds to the scalar field $X\equiv X^{ab}$, where $a,b$ are flavor indices.
The left- and right-handed current densities $J^\mu_L(x)$ and $J_R^\mu(x)$ at the boundary are respectively dual to the left-handed [right-handed] gauge potential fields $L^a_M$ and $R^a_M$ in the AdS bulk. 
These fields can be represented as $L_M = L_M^a t^a$ and $R_M = R_M^a t^a$, where $t^a$ are the generators of the special unitary group such that 2\,Tr$[t^at^b]=\delta^{ab}$. Thus, one can write the gauge field strength tensor as $F^L_{MN} = \partial_{[M} L_{N]} - i[L_M,L_N]$, and analogously for $F^R_{MN}$. 
Given these configurations, the action of the soft-wall (SW) model for holographic QCD can be written as
\begin{eqnarray}\label{g555}
        \mathcal{S}_\text{SW} &= \displaystyle{\int {\rm d}^5x \; e^{\phi} \sqrt{-g}\Tr\left\{\abs{D^M X}^2+V_C(\abs{X},\phi)-\frac{1}{4g_5^2}\left(F_L^2+F_R^2\right)\right\}},
\end{eqnarray}
with $F_L^2 = F_{MN}^L F^{LMN}$ and $F_R^2 = F_{MN}^R F^{RMN}$ \cite{Erlich:2005qh}.  
The covariant derivative is given by $D_MX = \partial_M X - i L_M X + i X R_M$ and $V_C$ is a scalar field potential that encompasses the interactions between $X$ and $\phi$ and will be discussed in more detail later on.
 The scalar field ${X}$ promotes a mapping to the operator $\langle\bar{q}_{R}q_{L}\rangle$ on the boundary, while the two gauge fields ${L}_M^a$ and ${R}_M^a$ correspond to $\langle\bar{q}_{{L}, {R}}\gamma_\mu t^aq_{{L},{R}}\rangle$ operator. The gauge
coupling $g_5$ in Eq. (\ref{g555}) can be obtained when the 2-point vector
correlation function matches the leading term entering the QCD operator product expansion involving 
 the vector 2-point function.  To implement it, one considers the 5-dimensional action governing solutions of the EOMs for the
vector field $V^M=A_L^M+A_R^M$. Subsequently, one can take two derivatives with respect to the boundary
sources to yield the vector 2-point function, given by $\int {\rm d}^{4}x\, e^{ipx}\, \langle J_{\mu}(x)
J_{\nu}(0)\rangle=\Pi_{V} (q^2)(q_{\mu}q_{\nu} -q^2 g_{\mu \nu})$. The near-boundary limit corresponding to large
scales in the boundary field theory yields  \cite{Erlich:2005qh} 
\begin{eqnarray}
    \Pi_V
(q^2)=-\frac{1}{2g_5^2}\ln{(q^2)}. \end{eqnarray} On the other hand, a quark bubble can be analyzed in QCD at high $q^2$, where QCD is
weakly-coupled due to asymptotic freedom, implying that  
\begin{eqnarray}
    \Pi_V
(q^2)=-\frac{N_c}{24\pi^2}\ln{(q^2)}\;.  \end{eqnarray}  Matching the holographic
model to QCD implies that $g_5^2=12\pi^2/N_c$.

The DhQCD model starts from the construction of a pure gluon system. This is implemented by coupling the dilaton field with five-dimensional background gravity, which breaks the conformal symmetry of the original AdS/CFT, as desired. This system is described by the 5-dimensional graviton-dilaton coupled action given by 
\begin{flalign}
    \mathcal{S}_\text{GD} &= \frac{1}{16\pi G_5} \int {\rm d}^5x \sqrt{-g} e^{-2\phi} \left(R+4\partial_M\phi \partial^M\phi - V(\phi)\right),
\end{flalign}
where $G_5$ is the 5-dimensional Newton's gravitational constant, $R$ denotes the Ricci scalar,  and $V(\phi)$ stands for  the dilaton potential.

The coupling between gluon dynamics and flavor dynamics to describe the complete dynamics of QCD is achieved through the expression \cite{Chen:2022pgo}:
\begin{flalign}
    \mathcal{S} = \mathcal{S}_\text{GD} + \frac{N_f}{N_c} \mathcal{S}_\text{SW}.
    \label{coupling}
\end{flalign}
Here, the ratio between the number of flavors $N_f$ and the number of colors $N_c$ represents a coupling constant. Thus, the two systems decouple if $N_f \ll N_c$. In this work,  $N_f=2$ and $N_c=3$ are considered.
Regarding the mesons here  analyzed formed only by up and down quarks, one can consider the isospin symmetry and treat all quark flavors as degenerate. Thus, the vacuum expectation value of the scalar field simplifies to $\langle X \rangle = (\upchi/2)I$, where the nonzero value $\upchi$ is responsible for the chiral symmetry breaking and $I$ is the identity matrix.

To write down the full action that determines the background dynamics of the fields in the vacuum, one assumes, in addition, that the gauge fields are perturbations above the vacuum and therefore are zero in the vacuum state.
Hence the complete action of DhQCD can be written using Eq. (\ref{coupling}) as
\begin{flalign}\label{act1}
    \mathcal{S} = \frac{1}{16\pi G_5} \int {\rm d}^5x\sqrt{g} \left[ e^{-2\phi} \bigg( R+4\partial_M\phi \partial^M\phi - V(\phi)  \bigg) 
    - \alpha e^{-\phi} \left( \frac{1}{2}\partial_M \upchi \partial^M \upchi + V_C(\upchi,\phi)
    \right)\right],
\end{flalign}
where the coupling constant between the two sectors is rewritten as $\alpha=16\pi G_5N_f/N_c$, under the assumption of unit AdS radius $L = 1$, for simplicity.
From the action (\ref{act1}) and the metric (\ref{adsmetric}), the equations of motion obtained are given by
\begin{subequations}
\begin{eqnarray}
    -A'' + {A'}^2 + \frac{2}{3}\phi'' - \frac{4}{3}A'\phi' - \frac{\alpha}{6}e^{\phi}{\upchi'}^2 &=&0,\\
    \upchi'' + (3A'-\phi')\upchi' - e^{2A}V_C &=& 0, \label{eqofmotionforx}\\
    \phi'' + (3A' - 2\phi')\phi' - \frac{3\alpha}{16}e^{\phi}{\upchi'}^2 - \frac{3}{8}e^{2A-\frac{4}{3}\phi} \partial_\phi \bigg(V(\phi) + \alpha e^{\frac{7}{3}\phi}V_{C,\upchi}\bigg) &=&0,
\end{eqnarray}
\end{subequations}
where the prime denotes a derivative with respect to the coordinate $z$ and $V_{C,\upchi}=\partial V_C/\partial\upchi$.
Therefore, it is sufficient to know the dilaton $\phi$ and the scalar field $\upchi$ to find the solutions for the potentials $V_\phi$ and $V_C$ by solving the equations of motion.

The expression for the scalar vacuum expectation value $\upchi$ is obtained by solving Eq. (\ref{eqofmotionforx}). When $\kappa=0$, the solution takes the form
\begin{flalign}
    \upchi(z) = c_1 z + c_2 z^3,
    \label{generalform}
\end{flalign}
and, through certain constraints, it is possible to associate the constants $c_1$ and $c_2$ with the quark mass $m_q$ and the chiral condensate $\sigma=\langle\bar{q}q\rangle$, respectively. Since the isospin symmetry holds, these parameters can be considered equal for the quarks involved.

However, in the standard soft-wall model, with $\phi(z)=\kappa^2z^2$ and $A(z)=\log(L/z)$ the solution has a more elaborate form \cite{Colangelo:2008us,Karch:2006pv},
\begin{flalign}
    \upchi(z) = \frac{m_q z}{L}\Gamma\left(\frac{3}{2}\right)U\left(\frac{1}{2},0,\kappa^2z^2\right),
    \label{smallz}
\end{flalign}
where $U$ is the Tricomi confluent hypergeometric function. In the small-$z$ limit, the solution (\ref{smallz}) acquires the following form:
\begin{flalign}
    \lim_{z\rightarrow 0}\,\upchi(z)\,=\, \frac{m_q}{L}z + \frac{m_q \kappa^2}{2L}z^3\left(1-2\gamma_E - 2\log(\kappa z) - \uppsi\left(\frac{3}{2}\right) \right)
\end{flalign}
where $\uppsi$ is the Euler function and $\gamma_E= \lim _{n\to \infty }\left(-\ln n+\sum _{k=1}^{n}{\frac {1}{k}}\right)\approx 0.577216$ is the Euler constant. By comparing this result with (\ref{generalform}), identifying the second term with the chiral condensate, we observe that the condensate now depends on the quark mass. This conclusion does not exist in QCD, and therefore, any model dual to it cannot include such a property, requiring some modifications.

Moreover, in the large-$z$  limit, the solution (\ref{smallz}) presents two possible forms: one is of the asymptotic limit  $\upchi\rightarrow m_q/\kappa$, which interferes with the larger mass meson resonances, generating non-linear trajectories. The other one  behaves as $\upchi\rightarrow\text{constant}$, which is the only solution that meets the model constraints  although it might suggest a restoration of chiral symmetry for large values of $n$, which is also not corroborated by QCD \cite{Gherghetta:2009ac}. 
Among the modifications to address these issues, Ref. \cite{Cherman:2008eh} proposes the solution in its asymptotic UV form given by:
\begin{flalign}
\lim_{z\rightarrow 0}    \upchi(z) = m_q \zeta z + \frac{\sigma}{\zeta} z^3,
    \label{uvsolution}
\end{flalign}
where $2\pi\zeta=\sqrt{N_c/N_f}$ is a normalization constant. As presented in Refs. \cite{Li:2012ay,Li:2013oda}, the asymptotic solution in the IR can be written as
\begin{flalign}
    \lim_{z\rightarrow \infty}\upchi(z)= \sqrt{\frac{8}{\alpha}}\kappa e^{-\phi(z)/2}.
    \label{irsolution}
\end{flalign}
Therefore, for $\upchi$ to assume the asymptotic forms (\ref{uvsolution}) and (\ref{irsolution}), it must have a general form as follows:
\begin{flalign}
    \upchi'(z) = \sqrt{\frac{8}{\alpha}}\kappa e^{-\phi(z)/2} \left(1+a_1e^{-\phi(z)} + a_2e^{-2\phi(z)}\right),
    \label{chi_condition}
\end{flalign}
with
\begin{flalign}
    a_1 &= -2 + \frac{5\sqrt{2\alpha}m_q\zeta}{8\kappa} + \frac{3\sqrt{2\alpha}\sigma}{4\zeta\kappa^3}, \\
     a_2 &= 1 - \frac{3\sqrt{2\alpha}m_q\zeta}{8\kappa} - \frac{3\sqrt{2\alpha}\sigma}{4\zeta\kappa^3}.
\end{flalign}

The potential $V_C$ can be explored to work on extensions of the soft-wall model. For results consistent with hadron spectroscopy experiments \cite{Bernardini:2018uuy}, it is sufficient to consider the potential in the form
\begin{flalign}
    V_C  = M_X^2 \abs{X}^2 = -3\abs{X}^2.
\end{flalign}
The 5-dimensional mass $M_X$ is obtained through its relation to the conformal scaling dimension of the 4-dimensional operator, which in the scalar case is given by the expression $M^2=\Delta(\Delta-4)$, with $\Delta=3$ for the $\bar{q}q$ operator. AdS/QCD posits  that fields in the AdS bulk are dual to the mesonic states on the AdS boundary. This duality can be precisely  implemented when one takes into account the conformal scaling  dimension $\Delta$, fixing the way the AdS bulk field scales at the AdS boundary. From the dual QCD point of view,  $\Delta$ plays the role of the scaling dimension underlying the operator responsible for creating mesons. The slopes for Regge trajectories describing light-flavor mesons, which is under scrutiny here, are  regulated by the dilaton energy scale $\kappa$, whereas $m_5^2$ stands for the AdS bulk mass of mesons.  Gauge/gravity asserts that scalar AdS bulk field  solutions, which are dual to an  $\mathcal{O}$  operator, in the boundary CFT scale as $z^{\Delta-4}$, in the UV limit, with 2-point correlation function reading  $
\langle\mathcal{O}(x)\,\mathcal{O}(\vec{0})\rangle \propto \left|x\right|^{-2\Delta}$ \cite{Witten:1998qj}. The perturbative calculations in QCD suggest a correction in the conformal dimension of the operators when quantum fluctuations are incorporated. With this correction in the form of $\Delta-\gamma(z)$, where $\gamma(z)$ is the anomalous dimension dependent on the energy scale $\sim 1/z$ \cite{Chen:2022pgo,Cherman:2008eh}. When the anomalous 5-dimension mass correction to the
scalar field is taken into account, the light-flavor meson  spectra and the pion
form factor can be simultaneously reconciled into a single setup.

It is effectively considered in the AdS side of the duality using a
mass term that depends on the holographical coordinate along the AdS bulk, for dual modes associated to the operator with anomalous dimensions. The AdS/CFT dictionary is used in bottom-up
models to dictate the bulk field content: a $p$-form 4-dimensional QFT operator
$\mathcal{O}$ with scaling dimension $\Delta$ corresponds to a $p$-form
bulk field with mass $m_{5}^2=(\Delta -p)(\Delta+p-4)$.  Conserved
currents in the QFT correspond to gauge fields in the bulk, with
$m_{5}^2=0$.  In bottom-up holographic models, it is assumed that this
dictionary remains valid even when one modifies the 5-dimensional background away
from AdS$_5$.  This property is not straightforward, as although the scaling dimension of operators is
well-defined in a CFT, in QCD the scaling dimensions of most operators
receive scale-dependent corrections proportional to the inverse of the ’t Hooft coupling constant $\lambda\equiv g^2 N_c$. In other words, most operators in QCD have scale-dependent anomalous dimensions  \cite{Cherman:2008eh}. The expression for the 5-dimensional mass is also corrected as:
\begin{flalign}
    M^2 = \big[\Delta -\gamma(z) \big] \big[ \Delta - \gamma(z) - 4\big].
    \label{masscorrected}
\end{flalign}
These corrections extend to the scalar field $\upchi$, which acquires the following asymptotic form:
\begin{eqnarray}\label{fas}
\lim_{z\rightarrow 0}    \upchi(z) = m_q \upzeta z^{1+\gamma(z)} + \frac{\sigma}{\upzeta} z^{3-\gamma(z)},
\end{eqnarray}
where $\upzeta = \zeta z^{\gamma(z)}$ was redefined to make the mass dimension of the terms on the right-hand side of Eq. (\ref{fas}) compatible.
Then, the condition (\ref{chi_condition}) with this insertion of the anomalous dimension effectively becomes
\begin{flalign}\label{fas1}
    \upchi'(z) = \sqrt{\frac{8}{\alpha}}\kappa e^{-\phi(z)/2} \left(1+a_1(z)e^{-\phi(z)} + a_2(z)e^{-2\phi(z)}\right),
\end{flalign}
with
\begin{eqnarray}\label{fas2}
    a_1(z) &=& -2 + \frac{5\sqrt{2\alpha}m_q e^2\upzeta}{8\kappa} + \frac{3\sqrt{2\alpha}\sigma }{4e^2\upzeta \kappa^3}, \\
     a_2(z) &=& 1 - \frac{3\sqrt{2\alpha}m_q e^2\upzeta}{8\kappa} - \frac{3\sqrt{2\alpha}\sigma }{4e^2\upzeta \kappa^3}, 
\end{eqnarray}
where the factor $e^2$ is added to ensure the dimension to recovered in the UV limit.

As shown in Ref. \cite{Gubser:2008yx}, it is possible to relate the dynamic evolution of the anomalous dimension to the beta function, expressing  the dependence of the coupling constant $g$ on the renormalization scale $\mu$,
\begin{flalign}
    \beta(g) = \mu\frac{dg}{d\mu},
\end{flalign}
effectively describing the running of the coupling constant. In perturbative QCD, the beta function is typically expanded in terms of the powers of the coupling constant, which can be calculated from the appropriate order loop corrections.
To fit the soft-wall model, one considers the 't Hooft coupling  and the renormalization scale $\mu$ is a function of the inverse of the $z$ coordinate so that one can  rewrite  the original expression of the beta function as
\begin{flalign}
    \beta(\lambda(z)) = -z\frac{d\lambda(z)}{dz}.
\end{flalign}
Following Ref. 
\cite{FolcoCapossoli:2016uns}, the beta function  reproduces the perturbative behavior of the UV regime in a one-loop approximation in the form $\beta(\lambda)\sim b_0\lambda^2$, with $b_0=\frac{1}{8\pi^2}(\frac{11}{3}-\frac{2}{9}N_f)$. To also reproduce the non-perturbative behavior of the IR regime, a small extension is imposed on the beta function expression, in the form 
\begin{flalign}
    \beta(\lambda) = -b_0 \lambda^2 \bigg(1-\frac{\lambda}{\lambda_0}\bigg)^2.
    \label{modified_beta}
\end{flalign}
Thus, $\lambda_0$, defined at a fixed point in the IR, becomes an input parameter in the model.
Also, based on the one-loop correction results, one can find that the anomalous dimension can be expressed in terms of the coupling constant as
\begin{flalign}
    \gamma = \frac{(N_c^2-1)}{32\pi^2 N_c^2}\lambda,
\end{flalign}
which leads to
\begin{flalign}\label{eqr}
    \frac{d\gamma}{dz} = -\frac{(N_c^2-1)}{32\pi^2 N_c^2 z}\beta(\lambda),
\end{flalign}
whose solution allows us to obtain the expression for $\gamma$ as a function of the beta function, as desired.
To investigate the effects of these corrections, Ref. \cite{Chen:2022pgo} considered four cases: case I does not account for the anomalous dimension correction, meaning that $\gamma=0$ in this case. 

Case II considers the correction in the form 
\begin{eqnarray}
    \gamma(z)=-\frac{3}{2}\left(e^{-\phi^2(z)/2}-1\right),
    \label{caseii}
\end{eqnarray} 
with concomitant modifications to the input parameters such that \begin{eqnarray}
m_q&\mapsto&m_qz^{\gamma(z)}\left[\frac{1}{2}(1+\gamma(z))+\ln(z)\gamma'(z)\right],\\
    \sigma&\mapsto&\sigma z^{-\gamma(z)}\left[\frac{1}{2}(3-\gamma(z))-\ln(z)\gamma'(z)\right],
    \end{eqnarray}
    to test the dominance of the non-perturbative regime over the UV limit.
    
    Case III uses the dynamic evolution of $\gamma(z)$ from the beta function (\ref{modified_beta}) and (\ref{eqr}), whose explicit expression is the inverse function 
    \begin{eqnarray}
      \gamma^{-1}(z)=  \frac{32 \pi ^2 b_0
   N_c^2 }{N_c^2-1}\left[\frac{1}{z}+\frac{1}{2} \ln\frac{1-z}{1+z}\right]\ln (z). 
   \label{caseiii}
    \end{eqnarray}
    
    Case IV uses a constant correction, making $\gamma(z)=\gamma$ an input parameter.    

The values for the input parameters considered will be $m_q=5$ MeV, $\sigma = (240\;\text{MeV})^3$, $\kappa=0.43$ GeV, and the Newton's constant is fixed as $G_5=3L^3/4=0.750$ \cite{Li:2012ay}. To solve for $\gamma(z)$ in terms of the beta function, the coupling at the IR fixed point, $\lambda_0$, is set to 1 and the initial condition used is $\lim_{z \to 1/M_Z}\lambda(z) = 0.1184$, where $M_Z$ is the $Z$ boson mass, frequently used as a mass scale in the analysis of the running of the QCD coupling constant.
For case IV, the constant value of the anomalous dimension is given by \cite{Chen:2022pgo}
\begin{eqnarray}
    \gamma=0.45.   
\end{eqnarray}
Initially, introducing the anomalous dimension applied only to the field $X$ dual to the operator $\bar{q}q$, one expects the consequences of the corrections to be immediate in the scalar sector described by the model. The scalar and pseudoscalar mesons can be introduced by the respective fields $s$ and $\uppi$ interpreted as perturbations in the vacuum so that the expression for $X$ becomes
\begin{flalign}
    X = \bigg( \frac{\upchi}{2} + s\bigg)e^{2i\uppi^at^a}.
\end{flalign}

Usually, vector and axial-vector mesons can be obtained by dividing the $L$ and $R$ fields into $V$ and $A$ fields. In the absence of these fields in the vacuum, these two types of mesons can be reproduced by the perturbations $v_\mu$ and $a_\mu$.
The mass spectrum of the mesons can be obtained from the equations of motion of the fields. From this, we denote $\uppi_n$ the function that describes the $\uppi$ pseudoscalar mesons, and $s_n$ the one that represents the $f_0$ scalar mesons, while $v_n$ describes the $\rho$ vector mesons and $a_n$ represents the $a_1$ axial-vector mesons.

After a Bogoliubov transformation, the equations of motion for each of these functions take the form of the following Schrödinger-like equations
\begin{subequations}
    \setlength{\abovedisplayskip}{2pt}
    \setlength{\belowdisplayskip}{2pt}
    \begin{alignat}{4}
       \bigg(-\partial^2_z + V_s\bigg)s_n &= m_n^2s_n,\\
    \bigg(-\partial^2_z + V_{\uppi,\varphi}\bigg)\uppi_n &= m_n^2 (\uppi_n - e^{A} \upchi \varphi_n),\\
    \bigg(-\partial^2_z + V_\varphi\bigg)\varphi_n &= g_5^2 e^{A} \chi (\uppi_n - e^{A} \upchi \varphi_n),\\
    \bigg(-\partial^2_z + V_v\bigg)v_n &= m_n^2 v_n,\\
    \bigg(-\partial^2_z + V_a\bigg)a_n &= m_n^2 a_n,
    \end{alignat}
    \label{schrodingerlike}
\end{subequations}
with the respective potentials 
\begin{subequations}
    \begin{eqnarray}
      V_s &=& \frac{1}{2}(3A'' - \phi'') + \frac{1}{4}(3A' - \phi')^2 +e^{2A}V_{C,\upchi\upchi},\\
    V_{\uppi,\varphi} &=& \frac{1}{2}\bigg( 3A'' - \phi'' + \frac{2\upchi''}{\upchi} - \frac{2{\upchi'}^2}{\upchi^2}\bigg) + \frac{1}{4}\bigg(3A'-\phi'+\frac{2\upchi'}{\upchi}\bigg)^2,\\
    V_\varphi &=& \frac{1}{2}(A''-\phi'') + \frac{1}{4}(A'-\phi')^2,\\
    V_v &=&\frac{1}{2}(A''-\phi'') + \frac{1}{4}(A'-\phi')^2,\\
    V_a &=& \frac{1}{2}(A''-\phi'') + \frac{1}{4}(A'-\phi')^2 + g_5^2 e^{2A}\upchi^2,
    \end{eqnarray}
    \label{schrodingerlikepotentials}
\end{subequations}
\!\!where one denotes $V_{C,\upchi\upchi}=\partial^2V_C/\partial\upchi^2$ and the pseudoscalar field couples with the longitudinal component of the axial-vector, $a_\mu^{\tiny{\|}}$. Thus one introduces an auxiliary field $\varphi$ and imposes $a_\mu^{\tiny{\|}}=\partial_\mu\varphi$.

Therefore, the $n = 1, 2, 3, \ldots$ solutions to the Schrödinger-like equations (\ref{schrodingerlike}) for the $\uppi_n, s_n, v_n$ and $a_n$ functions form the families of meson resonances. The mass spectra are calculated numerically using the previously presented input parameters, and the results are respectively shown in Tables \ref{table:spectra_pseudoscalar}, \ref{table:spectra_axial}, \ref{table:spectra_scalar}, and \ref{table:spectra_vector}, for each meson family analyzed.
 
\begin{table}[h]
\begin{center}
--------------  $\uppi$ pseudoscalar mesons mass spectrum  ---------------
\medbreak
\begin{tabular}{||c|c|c||c|c|c|c||}
\hline\hline
        $\quad n \quad$ &\; State \;& $M_\text{exp}$ (MeV) & $M_\text{I}$ (MeV) & $M_\text{II}$ (MeV) & $M_\text{III}$ (MeV) & $M_\text{IV}$ (MeV)   \\\hline\hline
\hline
        1 & $\uppi^\pm$      & $139.57039 \pm 0.00017$ & \clt{$140 \pm 0.000178$} & \clt{$146 \pm 0.000192$} & \clt{$158 \pm 0.000133$} & \clt{$109 \pm 0.000171$} \\ \hline
        2 & $\uppi(1300)$    & $1300 \pm 100$ & \clt{$1600 \pm 123$} & \clt{$1408 \pm 108$} & \clt{$1410 \pm 108$} & \clt{$1415 \pm 109$} \\ \hline
        3 & $\uppi(1800)$    & $1810^{+9}_{-11}$ & \clt{$1897\pm12$} & \clt{$1873\pm 11$} & \clt{$1821\pm11$} & \clt{$1841\pm11$} \\ \hline
        4* & $\uppi(2070)$    & $2070 \pm 35$ & \clt{$2116 \pm 36$} & \clt{$2123 \pm 36$} & \clt{$2070 \pm 35$} & \clt{$2096 \pm 36$} \\ \hline
        5* & $\uppi(2360)$    & $2360 \pm 25$ & \clt{$2299 \pm 24$} & \clt{$2313 \pm 25$} & \clt{$2264 \pm 24$} & \clt{$2289 \pm 24$} \\ \hline\hline
\end{tabular}
\caption{Experimental and predicted mass spectra for $\uppi$ pseudoscalar mesons. The experimental masses are taken from the up-to-date data in PDG \cite{ParticleDataGroup:2024cfk}, and the estimated masses follow the four cases of correction in the anomalous dimension: $\gamma=0$ in case I; $\gamma(z)=-\frac{3}{2}(e^{-\phi^2/2}-1)$ in case II; $\gamma(z)$ dynamically resolved from the running coupling in case III; and $\gamma=0.45$ in case IV. Meson states marked with the ``\,*\,'' are omitted from the summary table in the PDG.} 
\label{table:spectra_pseudoscalar}
\end{center}
\end{table}

\begin{table}[H]
\begin{center}
--------------  $a_1$ axial-vector mesons mass spectrum  ---------------
\medbreak
\begin{tabular}{||c|c|c||c|c|c|c||}
\hline\hline
        $\quad n \quad$ &\; State \;& $M_\text{exp}$ (MeV) & $M_\text{I}$ (MeV) & $M_\text{II}$ (MeV) & $M_\text{III}$ (MeV) & $M_\text{IV}$ (MeV)   \\\hline\hline \hline
        1 & $a_1(1260)$ & $1230 \pm 40$ & \clt{$1316 \pm 43$} & \clt{$1222 \pm 40$} & \clt{$1163 \pm 38$} & \clt{$1232 \pm 40$} \\ \hline
        2 & $a_1(1640)$ & $1655 \pm 16$ & \clt{$1735 \pm 17$} & \clt{$1676 \pm 16$} & \clt{$1649 \pm 16$} & \clt{$1707 \pm 17$} \\ \hline
        3* & $a_1(1930)$ & $1930\pm71$ & \clt{$1969\pm71$} & \clt{$1971\pm71$} & \clt{$1926\pm70$} & \clt{$1980\pm72$} \\ \hline
        4* & $a_1(2095)$ & $2096 \pm 41 \pm 81$ & \clt{$2163 \pm 126$} & \clt{$2178 \pm 127$} & \clt{$2132 \pm 124$} & \clt{$2181 \pm 127$} \\ \hline
        5* & $a_1(2270)$ & $2270\pm55$ & \clt{$2336\pm57$} & \clt{$2356\pm57$} & \clt{$2311\pm56$} & \clt{$2358\pm57$} \\ \hline\hline 
\end{tabular}
\caption{Experimental and predicted mass spectra for $a_1$ axial-vector mesons. The experimental masses are taken from the up-to-date data in PDG  \cite{ParticleDataGroup:2024cfk}, and the estimated masses follow the four cases of correction in the anomalous dimension: $\gamma=0$ in case I; $\gamma(z)=-\frac{3}{2}(e^{-\phi^2/2}-1)$ in case II; $\gamma(z)$ dynamically resolved from the running coupling in case III; and $\gamma=0.45$ in case IV. Meson states marked with ``\,*\,'' are omitted from the summary table in the PDG.} 
\label{table:spectra_axial}
\end{center}
\end{table}

\begin{table}[h]
\begin{center}
--------------  $f_0$ scalar mesons mass spectrum  ---------------
\medbreak
\begin{tabular}{||c|c|c||c|c|c|c||}
\hline\hline
        $\quad n \quad$ & State & $M_\text{exp}$ (MeV) & $M_\text{I}$ (MeV) & $M_\text{II}$ (MeV) & $M_\text{III}$ (MeV) & $M_\text{IV}$ (MeV)   \\\hline\hline
\hline
        1 & $f_0(500)$ & $550^{+250}_{-180}$ & \clt{$231\pm105$} & \clt{$488\pm222$} & \clt{$539\pm245$} & \clt{$609\pm277$} \\ \hline
        2 & $f_0(980)$ & $990 \pm 20$ & \clt{$1106 \pm 22$} & \clt{$1000\pm 20$} & \clt{$1043 \pm 21$} & \clt{$1040 \pm 21$} \\ \hline
        3 & $f_0(1370)$ & $1370^{+218}_{-156}$ & \clt{$1395\pm222$} & \clt{$1267\pm202$} & \clt{$1366\pm217$} & \clt{$1357\pm216$} \\ \hline
        4 & $f_0(1500)$ & $1522 \pm 25$ & \clt{$1632 \pm 27$} & \clt{$1591 \pm 26$} & \clt{$1620 \pm 27$} & \clt{$1608 \pm 26$} \\ \hline
        5 & $f_0(1770)$ & $1784\pm16$ & \clt{$1846\pm17$} & \clt{$1848\pm17$} & \clt{$1837\pm16$} & \clt{$1824\pm16$} \\ \hline
        6 & $f_0(2020)$ & $1982 \pm 3^{+54}_{-0}$ & \clt{$2039\pm59$} & \clt{$2063\pm59$} & \clt{$2030\pm58$} & \clt{$2016\pm58$} \\ \hline
        7* & $f_0(2100)$ & $2095\pm19$ & \clt{$2215\pm20$} & \clt{$2249\pm20$} & \clt{$2206\pm20$} & \clt{$2192\pm20$} \\ \hline
        8* & $f_0(2330)$ & $2330\pm153$ & \clt{$2376\pm156$} & \clt{$2410\pm158$} & \clt{$2367\pm155$} & \clt{$2355\pm155$} \\ \hline
        9* & $f_0(2470)$ & $2470 \pm 4^{+4}_{-6}$ & \clt{$2520\pm8$} & \clt{$2544\pm8$} & \clt{$2516\pm8$} & \clt{$2504\pm8$} \\ \hline\hline
\end{tabular}
\caption{Experimental and predicted mass spectra for $f_0$ scalar mesons. The experimental masses are taken from the up-to-date data in PDG \cite{ParticleDataGroup:2024cfk}, and the estimated masses follow the four cases of correction in the anomalous dimension: $\gamma=0$ in case I; $\gamma(z)=-\frac{3}{2}(e^{-\phi^2/2}-1)$ in case II; $\gamma(z)$ dynamically resolved from the running coupling in case III; and $\gamma=0.45$ in case IV. Meson states marked with ``\,*\,'' are omitted from the summary table in the PDG.} 
\label{table:spectra_scalar}
\end{center}
\end{table}

\begin{table}[H]
\begin{center}--------------  $\rho$ vector mesons mass spectrum ---------------\medbreak
\begin{tabular}{||c|c|c||c|c|c|c||}
\hline\hline
        $\quad n \quad$ & State & $M_\text{exp}$ (MeV) & $M_\text{I}$ (MeV) & $M_\text{II}$ (MeV) & $M_\text{III}$ (MeV) & $M_\text{IV}$ (MeV)   \\\hline\hline
\hline
        1 & $\rho(770)$ & $775.26 \pm 0.23$ & \clt{$771 \pm 0.22$} & \clt{$729 \pm 0.22$} & \clt{$737 \pm 0.22$} & \clt{$741 \pm 0.22$} \\ \hline
        2 & $\rho'(1450)$ & $1350 \pm 20^{+20}_{-30}$ & \clt{$1143\pm34$} & \clt{$1135\pm34$} & \clt{$1136\pm34$} & \clt{$1137\pm34$} \\ \hline
        3 & $\rho(1450)$ & $1465 \pm 25$ & \clt{$1431 \pm 24$} & \clt{$1423 \pm 24$} & \clt{$1425 \pm 24$} & \clt{$1426 \pm 24$} \\ \hline
        4 & $\rho(1700)$ & $1720 \pm 20$ & \clt{$1670 \pm 19$} & \clt{$1663 \pm 19$} & \clt{$1665 \pm 19$} & \clt{$1666 \pm 19$} \\ \hline
        5* & $\rho(1900)$ & $1900^{+51}_{-60}$ & \clt{$1878\pm50$} & \clt{$1873\pm50$} & \clt{$1874\pm50$} & \clt{$1875\pm50$} \\ \hline
        6* & $\rho(2150)$ & $2150\pm172$ & \clt{$2065\pm165$} & \clt{$2061\pm165$} & \clt{$2062\pm165$} & \clt{$2062\pm165$} \\ \hline
        7* & $\rho(2270)$ & $2270\pm60$ & \clt{$2237\pm59$} & \clt{$2234\pm59$} & \clt{$2234\pm59$} & \clt{$2235\pm59$} \\ \hline\hline
\end{tabular}
\caption{Experimental and predicted mass spectra for $\rho$ vector mesons. The experimental masses are taken from the up-to-date data in PDG  \cite{ParticleDataGroup:2024cfk}, and the estimated masses follow the four cases of correction in the anomalous dimension: $\gamma=0$ in case I; $\gamma(z)=-\frac{3}{2}(e^{-\phi^2/2}-1)$ in case II; $\gamma(z)$ dynamically resolved from the running coupling in case III; and $\gamma=0.45$ in case IV. Meson states marked with the ``\,*\,'' are omitted from the summary table in the PDG.} 
\label{table:spectra_vector}
\end{center}
\end{table}

\begin{figure}[H]
  \centering
  \begin{minipage}[b]{0.49\textwidth}
    \includegraphics[width=\textwidth]{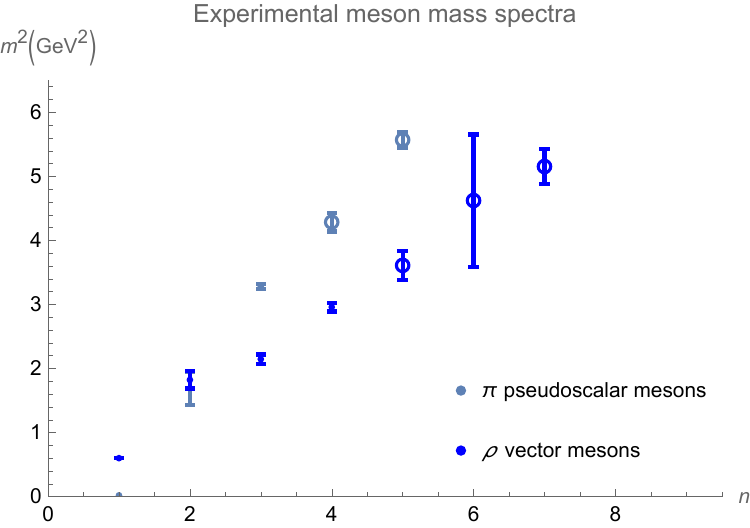}
    \caption{Mass spectra for the $\uppi$ and $\rho$ mesons from experimentally obtained masses as a function of the radial quantum number $n$.
    The grey dots represent the $\uppi^\pm$, $\uppi(1300)$, $\uppi(1800)$, $\uppi(2070)$, and $\uppi(2360)$ pseudoscalar mesons.
    The blue dots represent the $\rho(770)$, $\rho'(1450)$, $\rho(1450)$, $\rho(1700)$, $\rho(1900)$, $\rho(2150)$, and $\rho(2270)$ vector mesons. Meson states represented by open circles are omitted from the summary table in PDG.}
  \end{minipage}
  \hfill
  \begin{minipage}[b]{0.49\textwidth}
    \includegraphics[width=\textwidth]{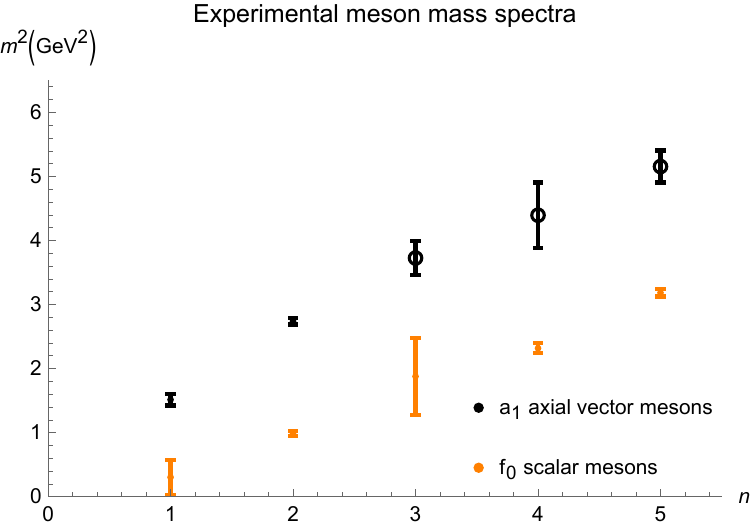}
    \caption{Mass spectra for the $a_1$ and $f_0$ mesons from experimentally obtained masses as a function of the radial quantum number $n$.   
    The black dots represent the $a_1(1260)$, $a_1(1640)$, $a_1(1930)$, $a_1(2095)$, and $a_1(2270)$ axial-vector mesons.
    The orange dots represent the $f_0(500)$, $f_0(980)$, $f_0(1370)$, $f_0(1500)$, $f_0(1770)$, $f_0(2020)$, $f_0(2100)$, $f_0(2330)$, and $f_0(2470)$ scalar mesons.
    States represented by open circles are omitted from the summary table in PDG.}
  \end{minipage}
\end{figure}
\noindent We will employ the DCE hereon to allow us to predict the mass spectra of heavier meson resonances in each one of the meson families, corresponding to resonances with $n$ beyond the data in PDG, which may correspond to either detected yet unidentified meson states or to still undetected meson resonances.

\section{DCE of light-flavor  mesons}
\label{section:dce}

The calculation of the DCE starts with the choice of the energy density as the localized function representing the meson families, which can be obtained from the timelike component of the stress-energy tensor:
\begin{flalign}
    \tau_{00} = \frac{2}{\sqrt{-g}}\left[ \frac{\partial(\sqrt{-g}\mathcal{L})}{\partial g^{00}} - \frac{\partial}{\partial x^\beta}\frac{\partial(\sqrt{-g}\mathcal{L})}{\partial\left(\frac{\partial g^{00}}{\partial x^\beta}\right)}
    \right],
\end{flalign}
where $\mathcal{L}$ is the Lagrangian density.
Thus, to obtain the CIMs of the system, it is only necessary to know its Lagrangian function, which can be read off the action (\ref{act1}), 
consisting of the integrand of the action. 
To create the configurational profile in momentum space, a Fourier transform of the tensor component is required,
\begin{flalign}\label{fou}
    \tau_{00}(q) = \frac{1}{(2\pi)^{k/2}} 
    \int_{-\infty}^{+\infty}
    \tau_{00}(x^M)e^{-iq_M x^M}\,{\sqrt{-g}}\,{\rm d}^kx,
\end{flalign}
where $q_M = (q_\mu, q_z)$ is the momentum associated with  the spatial components $x^M = (x^\mu,z)$. From this, it is possible to write the modal fraction for the DCE as:
\begin{flalign}
    \boldsymbol{\tau}_{00}(q) = \frac{\abs{\tau_{00}(q)}^2}{\displaystyle\int_{-\infty}^{+\infty} \abs{\tau_{00}({q})}^2 \,{\rm d}^k{q}}.
    \label{dce_modal}
\end{flalign}
The modal fraction is responsible for encoding the weight of each ${q}$ mode contribution to the total energy profile of the system. Thus, the DCE, which calculates the amount of information required to encode the energy density, is given by
\begin{flalign}
    \text{DCE} = -\int_{-\infty}^{+\infty}  \check{\boldsymbol{\tau}}_{00} ({q}) \ln\check{\boldsymbol{\tau}}_{00}({q}) {\rm d}^k{q},
    \label{dce}
\end{flalign}
where $\check{\boldsymbol{\tau}}_{00}({q})=\boldsymbol{\tau}_{00}({q})/\boldsymbol{\tau}_{00}^{\mathsf{max}}({q})$, and $\boldsymbol{\tau}_{00}^{\mathsf{max}}({q})$ denotes the maximum of $\boldsymbol{\tau}_{00}({q})$. The value $k=1$  will be fixed to estimate the DCE hereon,  according to the steps given by
the subsequent use of Eqs. (\ref{fou}) - (\ref{dce}), due to the codimension-1 AdS boundary of the extended AdS/QCD soft-wall model, as the dilaton, the warp factor,  and the $\upchi$ functions 
in the action (\ref{act1}) are $z$-dependent, only, and can be integrated along the AdS bulk, out of the AdS boundary. Therefore, for implementing numerical integration, Eq. (\ref{fou}) can be written, after 
a change of variables from $z\in[0,+\infty)$ to $\mathfrak{z}\in[0,1]$, as an integral over a finite interval $[0,1]$, instead of taking the entire real line as the integration interval\footnote{The integrals in Eqs. (\ref{fou}) - (\ref{dce}) are computed over the real line $(-\infty, +\infty)$. However, since $z$ represents the energy scale, such an interval consists of the non-negative 
 real axis $z\in[0,+\infty)$.}: 
\begin{eqnarray}\label{fou1}
    \tau_{00}(z) &=& \frac{1}{\sqrt{2\pi}} \int_{0}^{+\infty}\tau_{00}(z)e^{-iqz}dz\nonumber\\ &=& \frac{1}{\sqrt{2\pi}}\int _{0}^{1}\tau_{00}\left[\left({\frac{\mathfrak{z}}{1-\mathfrak{z}}}\right)\right]
    \exp\left[-iq\left({\frac{\mathfrak{z}}{1-\mathfrak{z}}}\right)
   \right]\frac{1}{\left(1-\mathfrak{z}\right)^{2}}\,{\rm d}\mathfrak{z}.
\end{eqnarray}
 Also, the integration over the real line in Eq. (\ref{dce}) can be rewritten within a finite integration interval as 
\begin{flalign}
    \text{DCE} = -\int_0^1 \check{\boldsymbol{\tau}}_{00}\left({\frac{\mathfrak{z}}{1-\mathfrak{z}}}\right)
    \ln\left\{\check{\boldsymbol{\tau}}_{00}\left({\frac{\mathfrak{z}}{1-\mathfrak{z}}}\right)\right\} 
    \frac{1}{\left(1-\mathfrak{z}\right)^{2}}\,{\rm d}\mathfrak{z},
    \label{dce1}
\end{flalign}
and solved by the Newton--Cotes quadrature method, using the composite iterated trapezoidal rule. We observe output convergence as long as the grid partition is progressively refined. Taking $1.9\times 10^6$ grid points keeps the numerical error below $10^{-8}$, and each DCE value  (\ref{dce1}) takes around 11.2 minutes on an 8 Core 4.8 GHz i9, under OsX Sonoma. As we will compute the DCE (\ref{dce1}) for 26 meson resonances in the $\uppi$, $a_1$, $f_0$, and $\rho$ families, in the next subsections, the numerical analysis, although repetitive, is straightforwardly feasible.

With the calculated DCE values, it is possible to express them both as a function of the radial quantum number $n$ and as a function of the squared mass of each mesonic state. In both cases, polynomial interpolation of the obtained points generates DCE trajectories similar to non-linear Regge trajectories. Combining both types of  DCE-Regge-like trajectories, respectively as a function of $n$ and the squared mass of meson resonances in each meson family, it is possible to extrapolate and estimate the masses of a new generation of light-flavor mesons. These newly estimated mesonic resonances may correspond to detected meson states that have not yet been identified to further meson states omitted from the summary table in PDG, which becomes one of the main phenomenological features of applying the DCE protocol in the context of AdS/QCD.

In addition to the ability to infer the next generation of light-flavor mesons, the DCE approximation is numerically calculated based on the experimentally obtained mass spectrum, making it phenomenologically robust and more accurate than obtaining the mass spectrum by solving the Schrödinger-like equations.

\subsection{$\uppi$ pseudoscalar mesons}
\label{subsection:pi}

    Based on the described protocol, the DCE for the pion family can be calculated. The values for the DCE of each meson resonance  are presented in Table \ref{table:DCE_pion}, considering all four analyzed cases I - IV.

    \begin{table}[H]
\centering
\begin{tabular}{||c|c|c|c|c|c||}
\hline\hline
\multirow{2}{*}{\vspace{-0.5em}\;$n$\;} & \multirow{2}{*}{\vspace{-0.5em}State} & \multicolumn{4}{c|}{DCE (nat)} \\ \cline{3-6} 
 &  & Case I & Case II & Case III & Case IV \\ \hline
1 & $\uppi^\pm$ & 4.5837 & 4.8583 & 5.1290 & 4.6988 \\ \hline
2 & \;$\uppi(1300)$\; & 6.5613 & 6.8907 & 7.0809 & 6.6787 \\ \hline
3 & $\uppi(1800)$ & 8.6321 & 8.9870 & 9.2464 & 8.8342 \\ \hline
4 & $\uppi(2070)$ & 10.1594 & 10.7347 & 10.9556 & 10.4175 \\ \hline
5 & $\uppi(2360)$ &\; 11.9119 \; &\; 12.4429 \;&\; 12.8191 \;&\; 12.1823 \;\\ \hline\hline
\end{tabular}
\caption{DCE of the radial resonances of the pion meson family.}
\label{table:DCE_pion}
\end{table}

\begin{figure}[H]
			\centering
			\includegraphics[width=10.5cm]{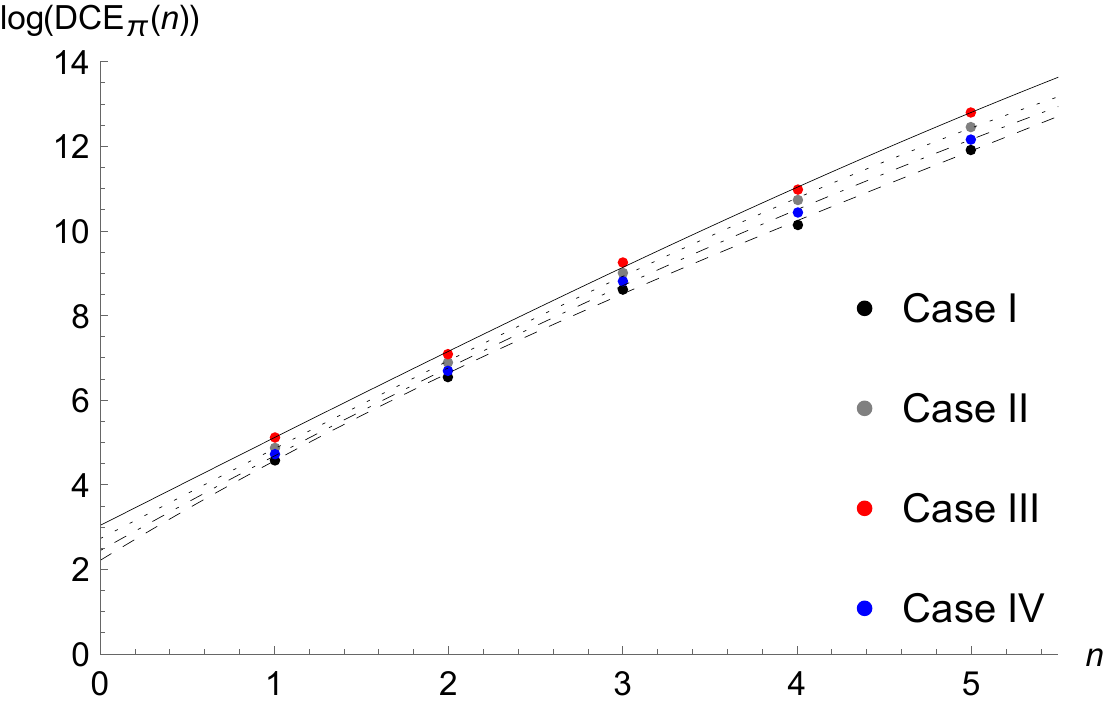}
			\caption{DCE of pion resonances. 
    The DCE is presented as a function of the $n$ radial excitation level, for  $n=1,\ldots,5$ (respectively corresponding to the $\uppi^\pm$, $\uppi(1300)$, $\uppi(1800)$, $\uppi(2070)$, and $\uppi(2360)$ resonances in PDG \cite{ParticleDataGroup:2024cfk}).  
    The pion resonances predicted by case I are depicted as black points; the ones estimated by case II are portrayed in grey, by case III in red, and by case IV in blue.
    The first forms of DCE-Regge-like trajectories are depicted following Eqs. (\ref{dce_pion_n1}) - (\ref{dce_pion_n4}), and are  respectively plotted as continuous, dotted, dot-dashed, and dashed lines.}
			\label{cen1}
		\end{figure}

	The first form of DCE-Regge-like trajectories regards the DCE as a function of the radial quantum number $n$ of pion resonances. Fig. \ref{cen1} illustrates the results, respectively for the cases I - IV,  whose quadratic polynomial  interpolation of data in Table \ref{table:spectra_pseudoscalar} generates the first types of DCE-Regge-like trajectory,
	\begin{subequations}
    \setlength{\abovedisplayskip}{2pt}
    \setlength{\belowdisplayskip}{2pt}
    \begin{alignat}{4}
        {{\rm DCE}}_{\uppi, \text{I}}(n) &= -7.109 \times 10^{-2} n^2 + 2.252 n + 2.395, \label{dce_pion_n1} \\
        {{\rm DCE}}_{\uppi, \text{II}}(n) &= -4.523 \times 10^{-2} n^2 + 2.197 n + 2.953, \label{dce_pion_n2} \\
        {{\rm DCE}}_{\uppi, \text{III}}(n) &= -7.161 \times 10^{-2} n^2 + 2.300 n + 2.449, \label{dce_pion_n3} \\
        {{\rm DCE}}_{\uppi, \text{IV}}(n) &= -7.121 \times 10^{-2} n^2 + 2.329 n + 2.580, \label{dce_pion_n4}
    \end{alignat}
    \label{dce_pion_n}
    \end{subequations}
Quadratic interpolation keeps the root mean square deviation (RMSD) within $10^{-3}$.
 
\begin{figure}[H]
			\centering
			\includegraphics[width=10.9cm]{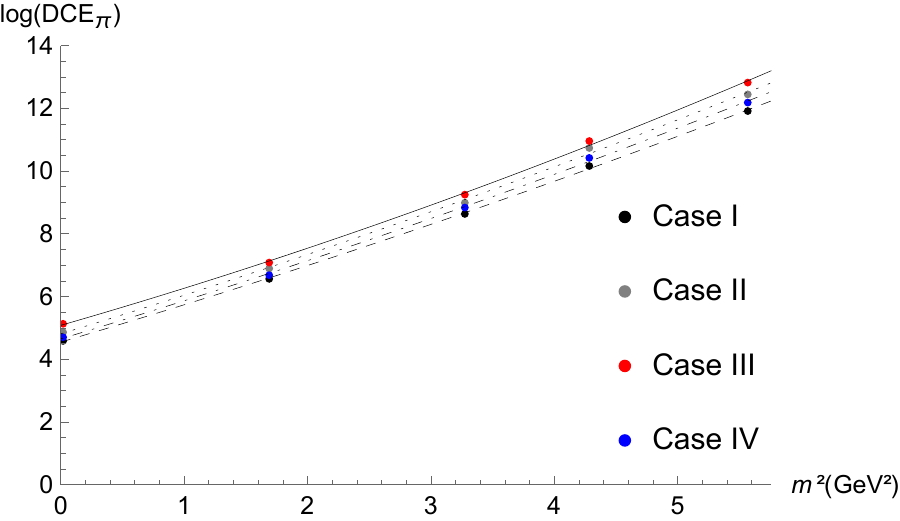}
			\caption{DCE of the pion  family  as a function of their squared mass, for  $n=1,\dots,5$ (respectively corresponding to the $\uppi^\pm$, $\uppi(1300)$, $\uppi(1800)$, $\uppi(2070)$, and $\uppi(2360)$ resonances in PDG \cite{ParticleDataGroup:2024cfk}).
            The second form of DCE-Regge-like trajectories is depicted following Eqs. (\ref{dce_pion_m1}) - (\ref{dce_pion_m4}). They are respectively plotted as continuous, dotted, dot-dashed, and dashed lines.}
			\label{cem11}
		\end{figure}

    The DCE of the pion family can also be expressed as a function of the experimental mass spectrum of the pions. Consequently, the second type of DCE-Regge-like trajectories is constructed from the experimentally observed mass spectrum for the pion radial resonances \cite{ParticleDataGroup:2024cfk}.
    By having the DCE of all pion radial resonances listed in Table \ref{table:DCE_pion}, we can plot the DCE against the  mass of each pion resonance (as shown in Table \ref{table:spectra_pseudoscalar}). The results are depicted in Fig. \ref{cem11}, where the interpolation method generates the second type of DCE-Regge trajectories presented in Eqs. (\ref{dce_pion_m}). 

  \begin{subequations}
    \setlength{\abovedisplayskip}{2pt}
    \setlength{\belowdisplayskip}{2pt}
    \begin{alignat}{4}
        {{\rm DCE}}_{\uppi, \text{I}}(m) &= 3.049 \times 10^{-2} m^4 + 1.159 m^2 + 4.5449, \label{dce_pion_m1} \\
        {{\rm DCE}}_{\uppi, \text{II}}(m) &= 3.343 \times 10^{-2} m^4 + 1.197 m^2 + 4.8094, \label{dce_pion_m2} \\
        {{\rm DCE}}_{\uppi, \text{III}}(m) &= 4.796 \times 10^{-2} m^4 + 1.133 m^2 + 5.0798, \label{dce_pion_m3} \\
        {{\rm DCE}}_{\uppi, \text{IV}}(m) &= 3.236 \times 10^{-2} m^4 + 1.182 m^2 + 4.6479. \label{dce_pion_m4}
    \end{alignat}
    \label{dce_pion_m}
    \end{subequations}
	within $0.023\%$ RMSD.
    
  Eqs. (\ref{dce_pion_n}, \ref{dce_pion_m}) encapsulate the main properties regarding the DCE of the pion meson family, making it possible to use them to estimate potential meson states for $n>5$. The procedure involves using the expressions for DCE as  functions of the radial quantum number $n$ (\ref{dce_pion_n1}) - \eqref{dce_pion_n4}, putting $n$ into Eqs. (\ref{dce_pion_n})  to obtain the corresponding DCE. These DCE values are then placed on the left-hand side of Eqs. (\ref{dce_pion_m1}) - \eqref{dce_pion_m4}, which relates it to the pion family mass spectrum,  respectively for each case I - IV. By solving this equation for $m$, the estimated mass for the meson  state at given $n$ can be obtained.
  Since these results are based solely on the DCE and the experimental mass spectrum of the pion meson family, their conclusions appear more realistic than the usual AdS/QCD methods, which involve solving the Schrödinger-like equation and obtaining the mass spectrum from its eigenvalues, as in Eqs. (\ref{schrodingerlike}).

  The first goal is to obtain the estimated mass for the $n=6$ resonance in the pion meson family, which we will call $\uppi_6^\star$. To implement it, the radial quantum number $n = 6$ is inserted into (\ref{dce_pion_n}), leading to the DCE values  equal to 13.3487 nat, 13.9882 nat, 14.5061 nat and 13.6728 nat, which correspond respectively to cases I - IV. It is worth emphasizing that the natural unit of information, symbolized by ``nat'', is used here as the fundamental unit of information entropy, based on the natural logarithm, alternatively to the base 2 logarithm defining the shannon (Sh). One nat corresponds to the  measure of the information underlying a random event with the probability of occurrence equaling $1/e$. The equivalence 1 nat $\approx 1/\ln 2$ shannons $\approx$ 1.44  Sh $\approx 1/\ln 10$ hartleys among the different units of information entropy hold.  Then, substituting these values for the DCE  into the left-hand side of (\ref{dce_pion_m}), we obtain algebraic quadratic equations that, when solved, give us the value of the pion resonance $m$ for each case I - IV, for $n=6$.

The same method can be applied to higher resonance meson  states. When $n=7$ is inserted into Eqs. (\ref{dce_pion_n1}) - (\ref{dce_pion_n4}), the resulting DCE values for cases I - IV are 14.6766 nat, 15.3910 nat, 16.115 nat, and 15.0421 nat, respectively. This allows us to solve Eqs.  (\ref{dce_pion_m1}) - (\ref{dce_pion_m4}) and obtain a mass value $m$ for $\uppi^\star_7$ in each case.
Similarly, $n=8$ generates DCE values equal to 17.0573 nat, 16.6514 nat, 17.6335 nat, and 16.2682 nat, permitting  us to calculate the masses for $\uppi^\star_8$ in the four cases.
The mass values obtained for these resonances in all four cases are summarized in Table \ref{table:pion_extended}.

      \begin{table}[H]
\begin{center}\medbreak
\begin{tabular}{||c|c|c||c|c|c|c||}
\hline\hline
        $\quad n \quad$ &\; State \;& $M_\text{exp}$ (MeV) & $M_\text{I}$ (MeV) & $M_\text{II}$ (MeV) & $M_\text{III}$ (MeV) & $M_\text{IV}$ (MeV)   \\\hline\hline
\hline
         1 & $\uppi^\pm$      & $139.57039 \pm 0.00017$ & \clt{$140 \pm 0.000178$} & \clt{$146 \pm 0.000192$} & \clt{$158 \pm 0.000133$} & \clt{$109 \pm 0.000171$} \\ \hline
        2 & $\uppi(1300)$    & $1300 \pm 100$ & \clt{$1600 \pm 123$} & \clt{$1408 \pm 108$} & \clt{$1410 \pm 108$} & \clt{$1415 \pm 109$} \\ \hline
        3 & $\uppi(1800)$    & $1810^{+9}_{-11}$ & \clt{$1897\pm12$} & \clt{$1873\pm 11$} & \clt{$1821\pm11$} & \clt{$1841\pm11$} \\ \hline
        4* & $\uppi(2070)$    & $2070 \pm 35$ & \clt{$2116 \pm 36$} & \clt{$2123 \pm 36$} & \clt{$2070 \pm 35$} & \clt{$2096 \pm 36$} \\ \hline
        5* & $\uppi(2360)$    & $2360 \pm 25$ & \clt{$2299 \pm 24$} & \clt{$2313 \pm 25$} & \clt{$2264 \pm 24$} & \clt{$2289 \pm 24$}  \\ \hline
        6$^\star$ &\; $\uppi^\star_6$ \;& ---     & \clt{2547 $\pm$ 56} & \clt{2547 $\pm$ 56} & \clt{2553 $\pm$ 56} & \clt{2546 $\pm$ 56} \\ \hline
        7$^\star$ &\; $\uppi^\star_7$ \;& ---     & \clt{2706 $\pm$ 59} & \clt{2708 $\pm$ 59} & \clt{2772 $\pm$ 61} & \clt{2706 $\pm$ 59} \\ \hline
        8$^\star$ &\; $\uppi^\star_8$ \;& ---     & \clt{2838 $\pm$ 62} & \clt{2841 $\pm$ 62} & \clt{2867 $\pm$ 63} & \clt{2838 $\pm$ 62}
     \\ \hline\hline
\end{tabular}
\caption{Table \ref{table:spectra_pseudoscalar} completed with the higher $n$  resonances  of the pion meson family. The four cases of anomalous mass correction are displayed in columns 4 - 7. The extrapolated masses for $n=6, 7, 8$ indicated with the ``$\,^\star\,$'' denote the values extrapolated by the concomitant use of the DCE-Regge-like trajectories (\ref{dce_pion_n}, \ref{dce_pion_m}), interpolating the experimental masses for $n=1,\dots,5$. The errors were propagated to the model predictions and displayed for the new pion resonances, for $n=6, 7, 8$.}
\label{table:pion_extended}
\end{center}
\end{table}

Considering the data available in the PDG, the $\uppi^\star_6$ and $\uppi^\star_8$ resonances have not found any possible matches. Only the $\uppi^\star_7$ resonance, except for case III, found a potential match with the further meson state $X(2680)$ in PDG \cite{ParticleDataGroup:2024cfk}. The predicted mass for this resonance in cases I, II, and IV are, respectively, 2706.92 $\pm$ 59 MeV, 2708.76 $\pm$ 59 MeV, and 2706.80 $\pm$ 59 MeV, which fall within the detected measured mass $2676\pm 27$ MeV for the $X(2680)$ meson.

\subsection{$a_1$ axial-vector mesons}
\label{subsection:a1}

A protocol analogous to the one already employed for pseudoscalar mesons represented by the pion family can now be adapted and applied to the $a_1$ meson family, instead. The DCE values for each $a_1$ meson  resonance, considering each of the four studied cases, are presented in Table \ref{table:DCE_a1}.
\begin{table}[H]
\centering
\begin{tabular}{||c|c|c|c|c|c||}
\hline\hline
\multirow{2}{*}{\vspace{-0.5em}\;$n$\;} & \multirow{2}{*}{\vspace{-0.5em}State} & \multicolumn{4}{c|}{DCE (nat)} \\ \cline{3-6} 
 &  & Case I & Case II & Case III & Case IV \\ \hline
1 & $a_1(1260)$ & 5.8237 & 6.4485 & 6.8135 & 6.0852 \\ \hline
2 & $a_1(1640)$ & 7.3888 & 8.2860 & 8.6997 & 7.7314 \\ \hline
3 & $a_1(1930)$ & 9.2468 & 10.1272 & 10.6694 & 9.4873 \\ \hline
4 & $a_1(2095)$ & 10.9983 & 12.2035 & 12.8673 & 11.3531 \\ \hline
5 & \;$a_1(2270)$\; &\; 12.6515 \;&\; 14.0436 \;&\; 14.8783 \;&\; 13.1214 \;\\ \hline\hline
\end{tabular}
\caption{DCE of the radial resonances of the $a_1$ meson family.}
\label{table:DCE_a1}
\end{table}

\begin{figure}[H]
			\centering
			\includegraphics[width=10.9cm]{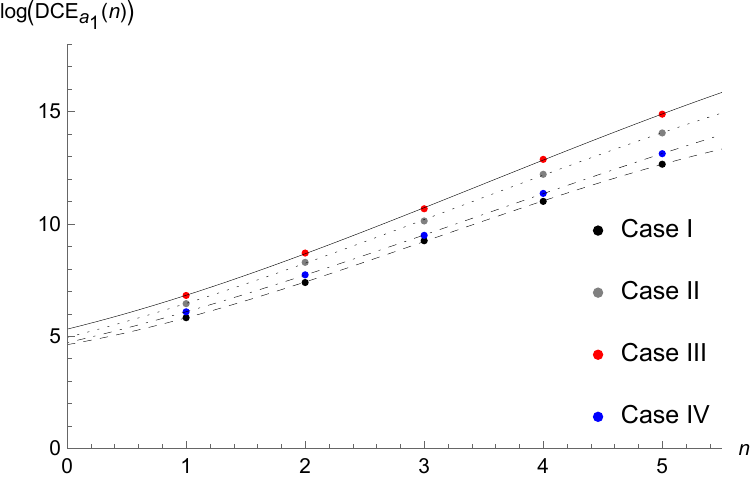}
			\caption{DCE of the $a_1$ resonances. The DCE is presented as a function of the $n$ radial excitation level, for  $n=1,\ldots,5$ (respectively corresponding to the $a_1(1260)$, $a_1(1640)$, $a_1(1930)$, $a_1(2095)$, and $a_1(2270)$ resonances in PDG \cite{ParticleDataGroup:2024cfk}).  
            The $a_1$ resonances predicted by case I are depicted as black points; the ones estimated by case II are portrayed in grey, by case III in red, and by case IV in blue.
            The first form of DCE-Regge-like trajectories is depicted following Eqs. (\ref{dce_a1_n1}) - 
 (\ref{dce_a1_n4}), and is respectively plotted as continuous, dotted, dot-dashed, and dashed lines.}
			\label{fig:dce_a1_n}
		\end{figure}

The first type of DCE-Regge-like trajectories is obtained by interpolating the DCE values as a function of the radial excitation level $n$ of the resonances from the values presented in Table \ref{table:spectra_axial}. The curves that interpolate these quantities are given by the following equations:
\begin{subequations}
    \setlength{\abovedisplayskip}{2pt}
    \setlength{\belowdisplayskip}{2pt}
    \begin{alignat}{4}
       \text{DCE}_{a_1,\text{I}}(n) &= 4.949 \times 10^{-3} n^2 + 1.697 n + 4.077, \label{dce_a1_n1} \\
        \text{DCE}_{a_1,\text{II}}(n) &= 3.414 \times 10^{-2} n^2 + 1.825 n + 4.935, \label{dce_a1_n2} \\
        \text{DCE}_{a_1,\text{III}}(n) &= 2.528 \times 10^{-2} n^2 + 1.618 n + 4.424, \label{dce_a1_n3} \\
        \text{DCE}_{a_1,\text{IV}}(n) &= 1.718 \times 10^{-2} n^2 + 1.808 n + 4.610, \label{dce_a1_n4}
    \end{alignat}
    \label{dce_a1_n}
\end{subequations}
within $0.002\%$ RMSD. The results are illustrated in Fig. \ref{fig:dce_a1_n}.

\begin{figure}[H]
			\centering
			\includegraphics[width=10.9cm]{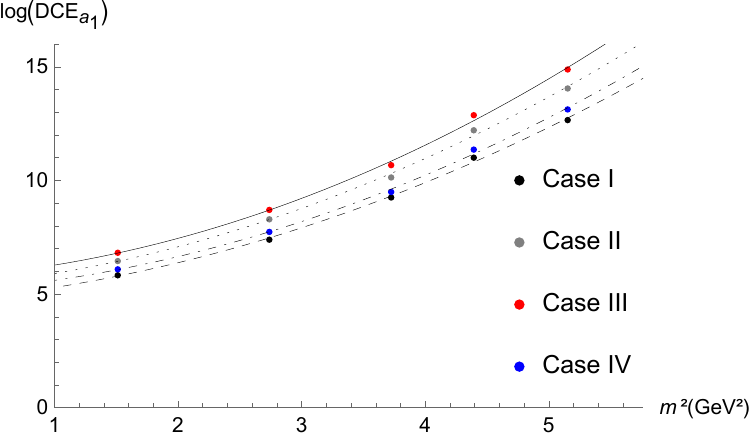}
			\caption{DCE of the $a_1$  family  as a function of their squared mass, for  $n=1,\dots,5$ (respectively corresponding to the $a_1(1260)$, $a_1(1640)$, $a_1(1930)$, $a_1(2095)$, and $a_1(2270)$ resonances in PDG \cite{ParticleDataGroup:2024cfk}).
            The second form of DCE-Regge-like trajectories is depicted following Eqs. (\ref{dce_a1_m1}) - (\ref{dce_a1_m4}), and is respectively plotted as continuous, dotted, dot-dashed, and dashed lines.}
			\label{fig:dce_a1_m}
		\end{figure}

As previously done, we can express the DCE values as a function of the mass of each $a_1$
  meson resonance. For this, we use the experimentally obtained values found in Table \ref{table:spectra_axial}. The plot of DCE versus the mass of each resonance can be seen in Fig. \ref{fig:dce_a1_m}, while the functions that interpolate these points, the second type of DCE-Regge-like trajectories, are given by:
\begin{subequations}
    \setlength{\abovedisplayskip}{2pt}
    \setlength{\belowdisplayskip}{2pt}
    \begin{alignat}{4}
        \text{DCE}_{a_1,\text{I}}(m) &= 2.235 \times 10^{-1} m^4 + 4.206 \times 10^{-1} m^2 + 4.6398, \label{dce_a1_m1} \\
        \text{DCE}_{a_1,\text{II}}(m) &= 2.553 \times 10^{-1} m^4 + 4.111 \times 10^{-1} m^2 + 5.2311, \label{dce_a1_m2} \\
        \text{DCE}_{a_1,\text{III}}(m) &= 2.906 \times 10^{-1} m^4 + 3.061 \times 10^{-1} m^2 + 5.6727, \label{dce_a1_m3} \\
        \text{DCE}_{a_1,\text{IV}}(m) &= 2.483 \times 10^{-1} m^4 + 3.011 \times 10^{-1} m^2 + 5.0481, \label{dce_a1_m4}
    \end{alignat}
    \label{dce_a1_m}
\end{subequations}
within $0.038\%$ RMSD.

Using Eqs. (\ref{dce_a1_n}, \ref{dce_a1_m}), which describe the main properties of the $a_1$ family, we can estimate meson states for $n>5$. The method involves substituting the desired $n$ value into the DCE($n$) expressions to find the corresponding DCE value. This value is then used in the mass relation  (\ref{dce_a1_m}) to solve for the estimated mass $m$.

The first heavier resonance in the $a_1$ family can be obtained for $n=6$ and will be denoted by  $({a_1})_6^\star$, the resonance for $n=6$. When this quantum number is applied in Eqs.  (\ref{dce_a1_n}), the DCE values obtained for cases I, II, III, and IV are, respectively, 14.4359 nat, 16.0743 nat, 17.1138 nat, and 15.0409 nat. By inserting these DCE values into the left side of Eqs. (\ref{dce_a1_m}), we obtain algebraic equations that can be solved to find the corresponding values of the mass spectrum $m$.

The same procedure is applied for $n=7$, which represents the state $({a_1})_7^\star$. For the four studied cases, the DCE values obtained from Eqs. (\ref{dce_a1_n}) are, respectively, given by: 16.1970 nat, 18.1054 nat, 19.3825 nat, and 16.9873 nat. This allows us to estimate the mass values for $({a_1})_7^\star$.
Finally, for $n=8$, repeating the procedures, we obtain the DCE values of 17.968 nat, 20.1708 nat, 21.7195 nat, and 18.9842 nat. The mass results obtained for $({a_1})_8^\star$, as well as for $({a_1})_6^\star$ and $({a_1})_7^\star$, are summarized in Table \ref{table:a1_extended}.

\begin{table}[H]
\begin{center}
\begin{tabular}{||c|c|c||c|c|c|c||}
\hline\hline
        $\quad n \quad$ &\; State \;& $M_\text{exp}$ (MeV) & $M_\text{I}$ (MeV) & $M_\text{II}$ (MeV) & $M_\text{III}$ (MeV) & $M_\text{IV}$ (MeV)   \\\hline\hline \hline
        1 & $a_1(1260)$ & $1230 \pm 40$ & \clt{$1316 \pm 43$} & \clt{$1222 \pm 40$} & \clt{$1163 \pm 38$} & \clt{$1232 \pm 40$} \\ \hline
        2 & $a_1(1640)$ & $1655 \pm 16$ & \clt{$1735 \pm 17$} & \clt{$1676 \pm 16$} & \clt{$1649 \pm 16$} & \clt{$1707 \pm 17$} \\ \hline
        3* & $a_1(1930)$ & $1930\pm71$ & \clt{$1969\pm71$} & \clt{$1971\pm71$} & \clt{$1926\pm70$} & \clt{$1980\pm72$} \\ \hline
        4* & $a_1(2095)$ & $2096 \pm 41 \pm 81$ & \clt{$2163 \pm 126$} & \clt{$2178 \pm 127$} & \clt{$2132 \pm 124$} & \clt{$2181 \pm 127$} \\ \hline
        5* & $a_1(2270)$ & $2270\pm55$ & \clt{$2336\pm57$} & \clt{$2356\pm57$} & \clt{$2311\pm56$} & \clt{$2358\pm57$} \\ \hline
        6$^\star$ &\; $({a_1})^\star_6$ \;& --- & \clt{2397 $\pm$ 70} & \clt{2400 $\pm$ 71} & \clt{2402 $\pm$ 71} & \clt{2401 $\pm$ 71} \\ \hline
        7$^\star$ &\; $({a_1})^\star_7$ \;& --- & \clt{2512 $\pm$ 74} & \clt{2518 $\pm$ 74} & \clt{2522 $\pm$ 74} & \clt{2520 $\pm$ 74} \\ \hline
        8$^\star$ &\; $({a_1})^\star_8$ \;& --- & \clt{2614 $\pm$ 77} & \clt{2624 $\pm$ 77} & \clt{2631 $\pm$ 77} & \clt{2628 $\pm$ 77}
 \\ \hline\hline
\end{tabular}
\caption{Table \ref{table:spectra_axial} completed with the higher $n$ resonances of the $a_1$ axial-vector family. The four cases of anomalous mass correction are displayed in columns 4 - 7. The extrapolated masses for $n=6, 7, 8$ indicated with a ``$^\star$'' denote the values extrapolated by the concomitant use of the DCE-Regge-like trajectories (\ref{dce_a1_n}, \ref{dce_a1_m}), interpolating the experimental masses for $n=1,\dots, 5$. The errors were propagated to the model predictions and displayed for the new $a_1$ resonances, for $n=6, 7, 8$.}
\label{table:a1_extended}
\end{center}
\end{table}

According to the data in PDG, the $({a_1})^\star_6$ axial-vector meson resonance has mass spectrum fitting, for all the four cases of the anomalous dimension, with the further meson state $X(2340)$, which has an experimental mass equal to $2340\pm20$ MeV.
The $({a_1})^\star_7$ resonance has not acquired any match with the further meson states cataloged in  PDG.
Finally, for the $({a_1})^\star_8$ axial-vector resonance, one can find possible matches for the further states $X(2600)$, $X(2632)$, and $X(2680)$ in all four cases.
These states have masses of $2618.3\pm2.0^{+16.3}_{-1.4}$ MeV, $2631.6 \pm 2.1$ MeV, and $2676 \pm 27$ MeV, respectively, covering all the predicted masses across the cases I, II, III, and IV.
 
\subsection{$f_0$ scalar mesons}
\label{subsection:f0}

The third group of particles analyzed will be the family of $f_0$ scalar mesons. The values obtained from applying the DCE protocol are listed in Table \ref{table:DCE_f0}.
\begin{table}[H]
\centering
\begin{tabular}{||c|c|c|c|c|c||}
\hline\hline
\multirow{2}{*}{\vspace{-0.5em}\;$n$ \;} & \multirow{2}{*}{\vspace{-0.5em}State} & \multicolumn{4}{c|}{DCE (nat)} \\ \cline{3-6} 
 &  & Case I & Case II & Case III & Case IV \\ \hline
1 & $f_0(500)$ & 1.1851 & 1.6410 & 1.8817 & 1.3796 \\ \hline
2 & $f_0(980)$ & 3.5514 & 3.8472 & 4.3234 & 3.7181 \\ \hline
3 & $f_0(1370)$ & 5.4049 & 5.7932 & 6.5052 & 5.6311 \\ \hline
4 & $f_0(1500)$ & 7.1802 & 7.7611 & 8.7170 & 7.5041 \\ \hline
5 & $f_0(1770)$ & 8.8281 & 9.4346 & 10.6263 & 9.0468 \\ \hline
6 & $f_0(2020)$ & 10.2014 & 10.9801 & 12.4143 & 10.7022 \\ \hline
7 & $f_0(2100)$ & 11.6711 & 12.5994 & 14.2711 & 12.2893 \\ \hline
8 & $f_0(2330)$ & 13.1114 & 14.0281 & 15.9391 & 13.6493 \\ \hline
9 &\; $f_0(2470)$ \;&\; 14.4653 \;&\; 15.5772 \;&\; 17.7300 \;&\; 15.1908 \; \\ \hline\hline
\end{tabular}
\caption{DCE of the radial resonances of the $f_0$ meson  family.}
\label{table:DCE_f0}
\end{table}

\begin{figure}[H]
			\centering
			\includegraphics[width=10.9cm]{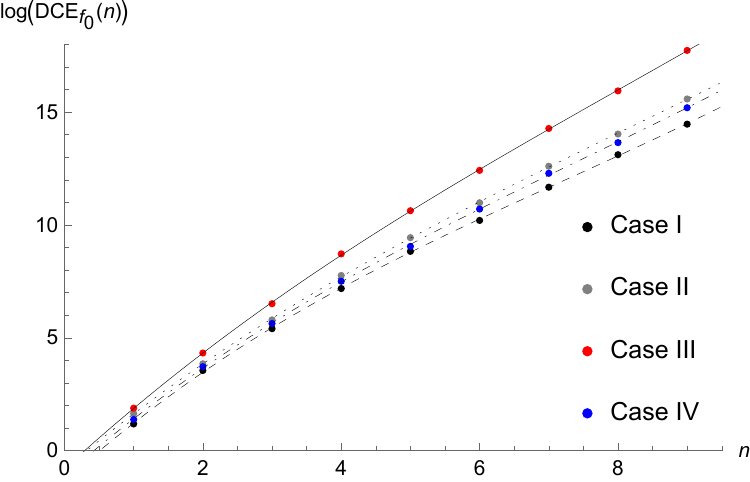}
			\caption{DCE of the $f_0$ resonances. The DCE is presented as a function of the $n$ radial excitation level, for  $n=1,\ldots,9$ (respectively corresponding to the $f_0(500)$, $f_0(980)$, $f_0(1370)$, $f_0(1500)$, $f_0(1770)$, $f_0(2020)$, $f_0(2100)$, $f_0(2330)$, and $f_0(2470)$ resonances in PDG \cite{ParticleDataGroup:2024cfk}).  
            The $f_0$ resonances predicted by case I are depicted as black points; the ones estimated by case II are portrayed in grey, by case III in red, and by case IV in blue.
            The first form of DCE-Regge-like trajectories is depicted following Eqs. (\ref{dce_f0_n1}) -  (\ref{dce_f0_n4}), and is respectively plotted as continuous, dotted, dot-dashed, and dashed lines.}
			\label{fig:dce_f0_n}
		\end{figure}

The first type of DCE-Regge-like trajectories is derived by interpolating the DCE values based on the radial excitation level $n$ of the resonances, using the data provided in Table \ref{table:spectra_scalar}. The results are illustrated in Fig. \ref{fig:dce_f0_n} and the curves that represent these interpolations are described by the following equations:
\begin{subequations}
  \setlength{\abovedisplayskip}{2pt}
    \setlength{\belowdisplayskip}{2pt}
    \begin{alignat}{4}
        \text{DCE}_{f_0,\text{I}}(n) &= 8.7296 \times 10^{-3} n^3 - 1.8928 \times 10^{-1} n^2 + 2.7579 n - 1.3609, \label{dce_f0_n1} \\
        \text{DCE}_{f_0,\text{II}}(n) &= 5.3675 \times 10^{-3} n^3 - 1.3152 \times 10^{-1} n^2 + 2.5684 n - 0.8109, \label{dce_f0_n2} \\
        \text{DCE}_{f_0,\text{III}}(n) &= 5.3554 \times 10^{-3} n^3 - 1.3131 \times 10^{-1} n^2 + 2.8062 n - 0.8090, \label{dce_f0_n3} \\
        \text{DCE}_{f_0,\text{IV}}(n) &= 7.2202 \times 10^{-3} n^3 - 1.6056 \times 10^{-1} n^2 + 2.6715 n - 1.1077, \label{dce_f0_n4}
    \end{alignat}
    \label{dce_f0_n}
\end{subequations}
	within $0.001\%$ RMSD. 

The DCE values can be expressed as a function of the experimental mass of each resonance, as shown in Table \ref{table:spectra_scalar}. These values are plotted against the mass values for the four cases of the anomalous dimension studied, in Fig. \ref{dce_f0_m}. The interpolating functions, which form the second type of DCE-Regge-like trajectories, are defined by the equations  
\begin{subequations}
 \setlength{\abovedisplayskip}{2pt}
 \setlength{\belowdisplayskip}{2pt}
    \begin{alignat}{4}
       \text{DCE}_{f_0,\text{I}}(m) &= 1.1377 \times 10^{-2} m^6 - 2.3773 \times 10^{-1} m^4 + 3.3474 m^2 + 0.2658, \label{dce_f0_m1} \\
        \text{DCE}_{f_0,\text{II}}(m) &= -2.3917 \times 10^{-4} m^6 - 1.1738 \times 10^{-1} m^4 + 3.4705 m^2 + 0.8764, \label{dce_f0_m2} \\
        \text{DCE}_{f_0,\text{III}}(m) &= 6.6807 \times 10^{-3} m^6 - 1.8313 \times 10^{-1} m^4 + 3.2675 m^2 + 0.4780, \label{dce_f0_m3} \\
        \text{DCE}_{f_0,\text{IV}}(m) &= 1.7533 \times 10^{-3} m^6 - 1.3350 \times 10^{-1} m^4 + 3.1689 m^2 + 0.7245. \label{dce_f0_m4}
    \end{alignat}
    \label{dce_f0_m}
\end{subequations}
	The RMSDs are 0.72\% for Eq. (\ref{dce_f0_m1}), 0.38\% for Eq. (\ref{dce_f0_m2}), 0.31\% for Eq. (\ref{dce_f0_m3}), and 0.28\% for Eq. (\ref{dce_f0_m4}). 

        \begin{figure}[H]
			\centering
			\includegraphics[width=10.9cm]{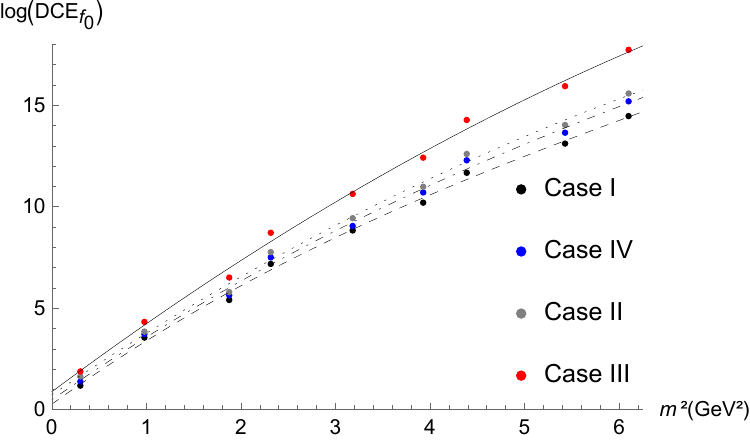}
			\caption{DCE of the $f_0$  family  as a function of their squared mass, for  $n=1,\dots,9$ (respectively corresponding to the $f_0(500)$, $f_0(980)$, $f_0(1370)$, $f_0(1500)$, $f_0(1770)$, $f_0(2020)$, $f_0(2100)$, $f_0(2330)$, and $f_0(2470)$ resonances in PDG \cite{ParticleDataGroup:2024cfk}).
            The second form of DCE-Regge-like trajectories is depicted following Eqs. (\ref{dce_f0_m1})  - (\ref{dce_f0_m4}), and is respectively plotted as continuous, dotted, dot-dashed, and dashed lines.}
			\label{fig:dce_f0_m}
		\end{figure}

    Eqs. (\ref{dce_f0_n}, \ref{dce_f0_m}) allow us to estimate the $f_0$ states for $n>9$. By substituting the desired $n$ value into the DCE expressions (\ref{dce_f0_n}), we obtain the corresponding DCE value for each case. This value is then used in the mass relation (\ref{dce_f0_m}) to solve for the mass $m$.
    For $n=10$, the resonance corresponding to the state $({f_0})^\star_{10}$, Eqs. (\ref{dce_f0_n}) yield the following DCE values for cases I, II, III, and IV: 16.0197 nat, 17.0886 nat, 19.4774 nat, and 16.7715 nat. By inserting these values into (\ref{dce_f0_m}) and solving the resulting algebraic equations for $m$, we obtain the mass solutions summarized in Table \ref{table:f0_extended}. 
The same procedure can be applied for $n=11$, corresponding to the state $({f_0})^\star_{11}$. In this case, the DCE values are 17.6922 nat, 18.6717 nat, 21.2987 nat, and 18.4611 nat, respectively for cases I - IV. Applying these values to Eqs. (\ref{dce_f0_m}) and solving for $m$, we obtain the masses, which are also summarized in Table \ref{table:f0_extended}.

\begin{table}[H]
\begin{center}
\begin{tabular}{||c|c|c||c|c|c|c||}
\hline\hline
        $\quad n \quad$ & State & $M_\text{exp}$ (MeV) & $M_\text{I}$ (MeV) & $M_\text{II}$ (MeV) & $M_\text{III}$ (MeV) & $M_\text{IV}$ (MeV)   \\\hline\hline
\hline
        1 & $f_0(500)$ & $550^{+250}_{-180}$ & \clt{$231\pm105$} & \clt{$488\pm222$} & \clt{$539\pm245$} & \clt{$609\pm277$} \\ \hline
        2 & $f_0(980)$ & $990 \pm 20$ & \clt{$1106 \pm 22$} & \clt{$1000\pm 20$} & \clt{$1043 \pm 21$} & \clt{$1040 \pm 21$} \\ \hline
        3 & $f_0(1370)$ & $1370^{+218}_{-156}$ & \clt{$1395\pm222$} & \clt{$1267\pm202$} & \clt{$1366\pm217$} & \clt{$1357\pm216$} \\ \hline
        4 & $f_0(1500)$ & $1522 \pm 25$ & \clt{$1632 \pm 27$} & \clt{$1591 \pm 26$} & \clt{$1620 \pm 27$} & \clt{$1608 \pm 26$} \\ \hline
        5 & $f_0(1770)$ & $1784\pm16$ & \clt{$1846\pm17$} & \clt{$1848\pm17$} & \clt{$1837\pm16$} & \clt{$1824\pm16$} \\ \hline
        6 & $f_0(2020)$ & $1982 \pm 3^{+54}_{-0}$ & \clt{$2039\pm59$} & \clt{$2063\pm59$} & \clt{$2030\pm58$} & \clt{$2016\pm58$} \\ \hline
        7* & $f_0(2100)$ & $2095\pm19$ & \clt{$2215\pm20$} & \clt{$2249\pm20$} & \clt{$2206\pm20$} & \clt{$2192\pm20$} \\ \hline
        8* & $f_0(2330)$ & $2330\pm153$ & \clt{$2376\pm156$} & \clt{$2410\pm158$} & \clt{$2367\pm155$} & \clt{$2355\pm155$} \\ \hline
        9* & $f_0(2470)$ & $2470 \pm 4^{+4}_{-6}$ & \clt{$2520\pm8$} & \clt{$2544\pm8$} & \clt{$2516\pm8$} & \clt{$2504\pm8$}  \\ \hline
        10$^\star$ &\; $({f_0})^\star_{10}$ \;& --- & \clt{2653 $\pm$ 192} & \clt{2660 $\pm$ 192} & \clt{2661 $\pm$ 192} & \clt{2657 $\pm$ 192} \\ \hline
        11$^\star$ &\; $({f_0})^\star_{11}$ \;& --- & \clt{2832 $\pm$ 204} & \clt{2860 $\pm$ 206} & \clt{2862 $\pm$ 206} & \clt{2844 $\pm$ 205} \\ \hline\hline
\end{tabular}
\caption{Table \ref{table:spectra_scalar} completed with the higher $n$  resonances  of the $f_0$ family. The four cases of anomalous mass correction are displayed in columns 4 - 7. The extrapolated masses for $n=10, 11$ indicated with a ``$^\star$'' denote the values extrapolated by the concomitant use of the DCE-Regge-like trajectories (\ref{dce_f0_n}, \ref{dce_f0_m}), interpolating the experimental masses for $n=1,\dots,9$. The errors were propagated to the model predictions and displayed for the new $f_0$ resonances, for $n=10, 11$.} 
\label{table:f0_extended}
\end{center}
\end{table}

The estimated masses for the $f_0^\star$ scalar meson resonances show a relatively large error range, allowing for a wider variety of possible matches. The estimated values for the $({f_0})^\star_{10}$ resonance cover all experimentally obtained mass ranges for the further meson states $X(2600)$, $X(2632)$, and $X(2680)$, which are, respectively, $2618.3\pm2.0^{+16.3}_{-1.4}$ MeV, $2631.6 \pm 2.1$ MeV, and $2676 \pm 27$ MeV.
As for the $({f_0})^\star_{11}$ scalar meson resonance, although it also has a large range in the predicted masses, it only covers the detected mass of the further meson state $X(2680)$, for all the four cases of the anomalous dimension. A small exception is case I, which, with a mass of $2832\pm204$, also lies within the range of correspondence for the $X(2632)$ state.

\subsection{$\rho$ vector mesons}
\label{subsection:rho}

Finally, the DCE protocol can be applied to the family of $\rho$ vector mesons. The results for the DCE are summarized in the Table \ref{table:DCE_rho}.
\begin{table}[H]
\centering
\begin{tabular}{||c|c|c|c|c|c||}
\hline\hline
\multirow{2}{*}{\vspace{-0.5em}\;$n$\;} & \multirow{2}{*}{\vspace{-0.5em}State} & \multicolumn{4}{c|}{DCE (nat)} \\ \cline{3-6} 
&  & Case I & Case II & Case III & Case IV \\ \hline
1 & $\rho(770)$ & 3.4904 & 3.9357 & 4.4100 & 3.6840 \\ \hline
2 & $\rho'(1450)$ & 5.7290 & 6.4184 & 6.8178 & 6.0210 \\ \hline
3 & $\rho(1450)$ & 7.6540 & 8.4414 & 8.7933 & 8.0035 \\ \hline
4 & $\rho(1700)$ & 10.3740 & 10.9784 & 11.3980 & 10.8100 \\ \hline
5 & $\rho(1900)$ & 12.0950 & 12.7642 & 13.4660 & 12.3370 \\ \hline
6 & $\rho(2150)$ & 13.5550 & 14.7020 & 15.3020 & 14.0280 \\ \hline
7 &\; $\rho(2270)$ \;&\; 15.0450\; &\; 16.2040 \;&\; 16.8760 \;&\; 15.4860\; \\ \hline\hline
\end{tabular}
\caption{DCE of the radial resonances of the $\rho$ meson family.}
\label{table:DCE_rho}
\end{table}

\begin{figure}[H]
			\centering
			\includegraphics[width=10.9cm]{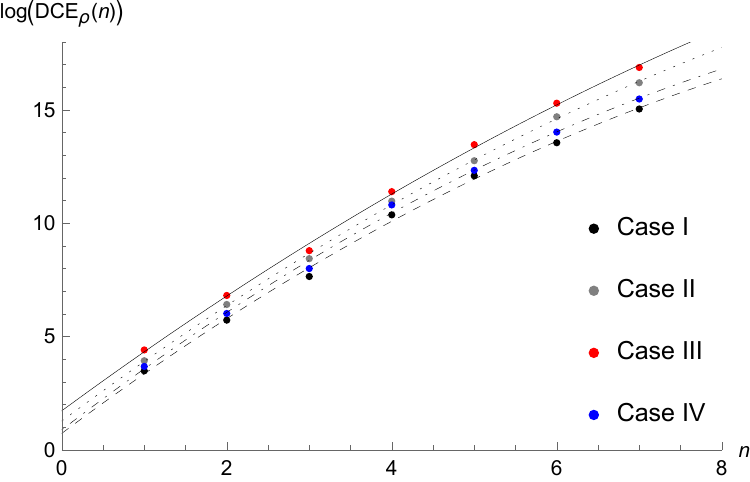}
			\caption{DCE of the $\rho$ resonances. The DCE is presented as a function of the $n$ radial excitation level, for  $n=1,\ldots,7$ (respectively corresponding to the $\rho(770)$, $\rho'(1450)$, $\rho(1450)$, $\rho(1700)$, $\rho(1900)$, $\rho(2150)$, and $\rho(2270)$ resonances in PDG \cite{ParticleDataGroup:2024cfk}).  
    The $\rho$ resonances predicted by case I are depicted as black points; the ones estimated by case II are portrayed in grey, by case III in red, and by case IV in blue.
    The first form of DCE-Regge-like trajectories is depicted following Eqs. (\ref{dce_rho_n1}) - (\ref{dce_rho_n4}), and is respectively plotted as continuous, dotted, dot-dashed, and dashed lines.}
			\label{fig:dce_rho_n}
		\end{figure}

The plot of the DCE values as a function of the radial quantum number $n$ is illustrated in Fig. \ref{fig:dce_rho_n}. Interpolating these values  we obtain the first type of DCE-Regge-like trajectories, which are described by Eqs. (\ref{dce_rho_n}).
\begin{subequations}
 \setlength{\abovedisplayskip}{2pt}
 \setlength{\belowdisplayskip}{2pt}
     \begin{alignat}{4}
        \text{DCE}_{\rho,\text{I}}(n) &= -0.0960 n^2 + 2.7240 n + 0.7306, \label{dce_rho_n1} \\
        \text{DCE}_{\rho,\text{II}}(n) &= -0.0813 n^2 + 2.7113 n + 1.2736, \label{dce_rho_n2} \\
        \text{DCE}_{\rho,\text{III}}(n) &= -0.0707 n^2 + 2.6743 n + 1.7269, \label{dce_rho_n3} \\
        \text{DCE}_{\rho,\text{IV}}(n) &= -0.1004 n^2 + 2.7939 n + 0.8848, \label{dce_rho_n4}
    \end{alignat}
    \label{dce_rho_n}
\end{subequations}
within $0.002\%$ RMSD.

\begin{figure}[H]
			\centering
			\includegraphics[width=10.9cm]{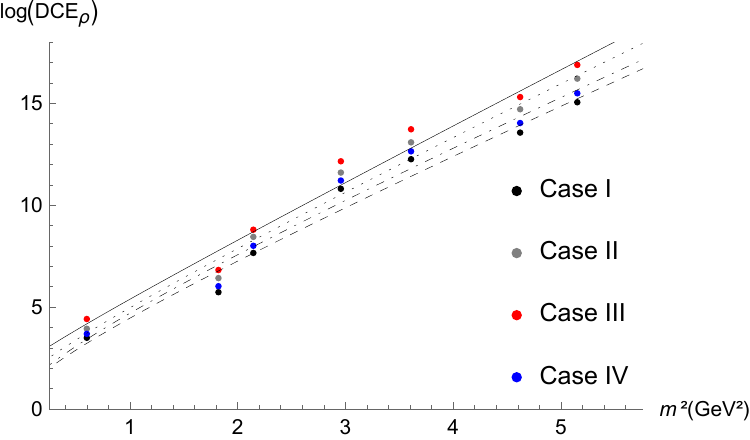}
			\caption{DCE of the $\rho$  family  as a function of their squared mass, for  $n=1,\dots,7$ (respectively corresponding to the $\rho(770)$, $\rho'(1450)$, $\rho(1450)$, $\rho(1700)$, $\rho(1900)$, $\rho(2150)$, and $\rho(2270)$ resonances in PDG \cite{ParticleDataGroup:2024cfk}).
The second form of DCE-Regge-like trajectories is depicted following Eqs. (\ref{dce_rho_m1}) - 
 (\ref{dce_rho_m4}), and is respectively plotted as continuous, dotted, dot-dashed, and dashed lines.}
			\label{fig:dce_rho_m}
		\end{figure}

DCE values can also be represented as a function of the experimental mass of each resonance, which are presented in Table \ref{table:spectra_vector}. The DCE values are plotted alongside the mass values for the four examined cases in Fig. \ref{fig:dce_rho_m}. The functions that interpolate these data points, forming the second type of DCE-Regge-like trajectories, are defined by Eqs.  (\ref{dce_rho_m}).
\begin{subequations}
\setlength{\abovedisplayskip}{2pt}
    \setlength{\belowdisplayskip}{2pt}
    \begin{alignat}{4}
        \text{DCE}_{\rho,\text{I}}(m) &= -1.6132 \times 10^{-1} m^6 + 1.2135 m^4 + 3.3776 \times 10^{-1} m^2 + 2.7424, \label{dce_rho_m1} \\
        \text{DCE}_{\rho,\text{II}}(m) &= -1.6831 \times 10^{-1} m^6 + 1.2868 m^4 + 3.2691 \times 10^{-1} m^2 + 3.6457, \label{dce_rho_m2} \\
        \text{DCE}_{\rho,\text{III}}(m) &= -1.6161 \times 10^{-1} m^6 + 1.2037 m^4 + 4.5905 \times 10^{-1} m^2 + 2.8653, \label{dce_rho_m3} \\
        \text{DCE}_{\rho,\text{IV}}(m) &= -1.4731 \times 10^{-1} m^6 + 1.0957 m^4 + 7.6060 \times 10^{-1} m^2 + 2.9784. \label{dce_rho_m4}
\end{alignat}
    \label{dce_rho_m}
\end{subequations}
within $1.52\%$ RMSD. 

To estimate the $\rho$ states for $n>7$, we use Eqs. (\ref{dce_rho_n}, \ref{dce_rho_m}).
The procedure, as before, involves inserting the value of $n$ into the expressions for the DCE in (\ref{dce_rho_n}). The resulting DCE values are then substituted into the left side of Eqs.  (\ref{dce_rho_m}), which are solved to obtain the mass values.

    For the resonance that we will call ${\rho}^\star_8$, substitute $n=8$ in Eqs. (\ref{dce_rho_n}) to obtain the DCE values for each of the four cases. The results for cases I, II, III, and IV are, respectively, 16.1674 nat, 17.4496 nat, 18.2374 nat, and 16.5997 nat. With these values and using Eqs. (\ref{dce_rho_m}), we obtain the solutions for the mass of this resonance in the four cases.
    The same can be done with $n=9$, resulting in the DCE values for the resonance ${\rho}^\star_9$ in the four cases: 17.0310 nat, 18.4236 nat, 19.3162 nat, and 17.4322 nat. The mass values obtained with these quantities are expressed in Table \ref{table:rho_extended}, as well as the masses of ${\rho}^\star_8$.
 
\begin{table}[H]
\begin{center}
\begin{tabular}{||c|c|c||c|c|c|c||}
\hline\hline
        $\quad n \quad$ & State & $M_\text{exp}$ (MeV) & $M_\text{I}$ (MeV) & $M_\text{II}$ (MeV) & $M_\text{III}$ (MeV) & $M_\text{IV}$ (MeV)   \\\hline\hline
\hline
        1 & $\rho(770)$ & $775.26 \pm 0.23$ & \clt{$771 \pm 0.22$} & \clt{$729 \pm 0.22$} & \clt{$737 \pm 0.22$} & \clt{$741 \pm 0.22$} \\ \hline
        2 & $\rho'(1450)$ & $1350 \pm 20^{+20}_{-30}$ & \clt{$1143\pm34$} & \clt{$1135\pm34$} & \clt{$1136\pm34$} & \clt{$1137\pm34$} \\ \hline
        3 & $\rho(1450)$ & $1465 \pm 25$ & \clt{$1431 \pm 24$} & \clt{$1423 \pm 24$} & \clt{$1425 \pm 24$} & \clt{$1426 \pm 24$} \\ \hline
        4 & $\rho(1700)$ & $1720 \pm 20$ & \clt{$1670 \pm 19$} & \clt{$1663 \pm 19$} & \clt{$1665 \pm 19$} & \clt{$1666 \pm 19$} \\ \hline
        5* & $\rho(1900)$ & $1900^{+51}_{-60}$ & \clt{$1878\pm50$} & \clt{$1873\pm50$} & \clt{$1874\pm50$} & \clt{$1875\pm50$} \\ \hline
        6* & $\rho(2150)$ & $2150\pm172$ & \clt{$2065\pm165$} & \clt{$2061\pm165$} & \clt{$2062\pm165$} & \clt{$2062\pm165$} \\ \hline
        7* & $\rho(2270)$ & $2270\pm60$ & \clt{$2237\pm59$} & \clt{$2234\pm59$} & \clt{$2234\pm59$} & \clt{$2235\pm59$}   \\ \hline
        8$^\star$ &\;$\rho^\star_8$ \;& --- & \clt{2315 $\pm$ 67} & \clt{2347 $\pm$ 68} & \clt{2336 $\pm$ 68} & \clt{2314 $\pm$ 67} \\ \hline
        9$^\star$ &\;$\rho^\star_9$ \;& --- & \clt{2341 $\pm$ 68} & \clt{2378 $\pm$ 69} & \clt{2368 $\pm$ 68} & \clt{2339 $\pm$ 68} \\ \hline\hline
\end{tabular}
\caption{Table \ref{table:spectra_vector} completed with the higher $n$  resonances  of the $\rho$ family. The four cases of anomalous mass correction are displayed in columns 4 - 7. The extrapolated masses for $n=8,9$ indicated with a ``$^\star$'' denote the values extrapolated by the concomitant use of the DCE-Regge-like trajectories (\ref{dce_rho_n}, \ref{dce_rho_m}), interpolating the experimental masses for $n=1,\dots, 7$. The errors were propagated to the model predictions and displayed for the new $\rho$ resonances, for $n=8, 9$.}
\label{table:rho_extended}
\end{center}
\end{table}

The mass values for the resonances ${\rho}^\star_8$ and ${\rho}^\star_9$ were surprisingly close, so all representatives of both states fall within the margin of error of the same further mesonic state in PDG, $X(2340)$. This state has an experimentally detected mass of $2340\pm20$ MeV, which lies within the predicted ranges for the four cases of both the ${\rho}^\star_8$ and ${\rho}^\star_9$ states, as shown in Table \ref{table:rho_extended}.
 
\section{Concluding remarks and perspectives}
\label{section:conclusion}
In this work, the DhQCD model was employed, which extends the description of the soft-wall model by adding gluon dynamics, complementing the flavor dynamics present in the original model. In addition to this formulation, the emergence of an anomalous dimension was considered, modifying the conformal dimension of the operators due to quantum fluctuations. As a consequence, a correction  to the 5-dimensional mass of the operators set in  (\ref{masscorrected}).
This correction was analyzed in four different cases: in case I, the correction was considered to be zero; in case II, the correction follows the expression (\ref{caseii}); in case III, the correction is taken to be dependent on the beta function (\ref{caseiii}); and in case IV, a constant correction is used. By considering these four cases, the possibility of finding correspondences among the further meson states is increased.

The implementation of DCE alongside AdS/QCD models has proven to be highly successful in hadron spectroscopy. By using the DCE method, two types of Regge-like trajectories can be obtained. The first one related the DCE as a function of the radial excitation quantum number $n$, whereas the other one established the DCE as a function of the squared mass of each resonance.
From this, it was possible to calculate the DCE for resonances beyond those identified by using larger values of $n$, thereby predicting the DCE for higher states. These values were then applied to the DCE expressions as a function of the squared meson masses, allowing for estimates of the masses for the next generation of meson resonances. The extrapolated masses were then compared with the mass values of experimentally detected -- but not yet identified -- states listed in the PDG, leading to possible correspondences.
In this context, four families of mesons were analyzed: the $\uppi$ pseudoscalar mesons, the $f_0$ scalar mesons, the $\rho$ vector mesons, and the $a_1$ axial-vector mesons. These particles have been extensively detected and studied, including in the context of DCE, so that their set of possible states appears to be complete. Thus, this work aimed to investigate the possibility of the existence of new, yet unidentified states of these mesons using the DhQCD model and the anomalous dimension correction as a different approach.

In Section \ref{subsection:pi}, the DCE values of the $\uppi$ pseudoscalar mesons were calculated for the four cases and presented in Table \ref{table:DCE_pion}. The interpolation of these values led to the curves for the DCE both as a function of the quantum number $n$, given by Eqs.  (\ref{dce_pion_n}), and as a function of the squared mass in Eqs. (\ref{dce_pion_m}), for each resonance corresponding to the first five known detected states. The extrapolated mass values for the resonances $n=6, 7, 8$ for the four cases are presented in Table \ref{table:pion_extended}.
The estimated masses for the $\uppi^\star_7$ state in cases I, II, and IV found a possible match with the further meson state $X(2680)$ found in PDG \cite{ParticleDataGroup:2024cfk}.

The DCE values of the $a_1$ axial-vector mesons are presented in Table \ref{table:DCE_a1} of Section \ref{subsection:a1}. The first five resonances were used for interpolation to obtain the Regge-like trajectories (\ref{dce_a1_n}, \ref{dce_a1_m}), and the extrapolated values were used to estimate the mass spectrum of the $a_1$ axial-vector meson states $n=6,7,8$ (presented in Table \ref{table:a1_extended}).
The $(a_1)^\star_6$ resonance found a possible match with the further meson state $X(2340)$ in all four cases of anomalous mass. This is the only instance among the results in this paper where the four cases of extrapolated mass for a resonance found a match exclusively in a single further state.
The other state that found possible matches was $(a_1)^\star_8$, where all the four cases coincided with the detected mass of the $X(2600)$, $X(2632)$, and $X(2680)$ meson states in PDG \cite{ParticleDataGroup:2024cfk}. Additional experiments may provide greater clarity in distinguishing among these results.

In Sec. \ref{subsection:f0}, the DCE of the $f_0$ scalar mesons are compiled in Table \ref{table:DCE_f0}. The Regge-like DCE trajectories (\ref{dce_f0_n}, \ref{dce_f0_m}) were obtained by interpolating nine resonances, and the extrapolated values for $n=10,11$ from the curves are presented in Table \ref{table:f0_extended}. 
The results we obtained have a large margin of error extrapolated from the experimentally detected values, so that the same resonances found more than one possible match among the further meson 
 states. The estimated mass for the $(f_0)^\star_{10}$ scalar meson resonance falls within the detected range for the meson 
 states $X(2600)$, $X(2632)$, and $X(2680)$ in PDG \cite{ParticleDataGroup:2024cfk}, for the four cases of anomalous dimension. For the $(f_0)^\star_{11}$ scalar meson resonance, two possible matches were found in case I -- the meson states states $X(2632)$ and $X(2680)$ -- whereas the remaining cases only coincide with the detected average mass of the $X(2680)$ meson state \cite{ParticleDataGroup:2024cfk}. 
 
In Sec. \ref{subsection:rho}, the $\rho$ vector mesons have their DCE values calculated and summarized in Table \ref{table:DCE_rho} for the seven first resonances in the PDG, with the Regge-like trajectories obtained from the interpolation of this data presented in Eqs.  (\ref{dce_rho_n}, \ref{dce_rho_m}). The extrapolated values, presented in Table \ref{table:rho_extended}, were calculated for the resonances $n=8,9$. The estimated masses for both resonances were very close to each other in all cases so that both $\rho^\star_8$ and $\rho^\star_9$ found a possible correspondence in the same further meson state from the PDG, $X(2340)$. 
Future runs of experiments can shed light on these identifications. 

The DCE method presents several considerable advantages compared to other methods derived from the soft-wall AdS/QCD model, such as being more succinct than its more widely known alternative, which involves solving Schrödinger-like equations (\ref{schrodingerlike}) to obtain the mass spectrum from the eigenvalues. Besides, the DCE takes into account experimental values of meson resonances in PDG to compute the mass spectrum of heavier meson resonances. It makes the DCE-based AdS/QCD hybrid method more robust, from the phenomenological point of view. 
The analysis of four different cases of anomalous dimension correction for four families of mesons was effective in reinforcing the property of DCE as a measure of configurational stability. As shown in Tables \ref{table:DCE_pion}, \ref{table:DCE_a1}, \ref{table:DCE_f0}, and \ref{table:DCE_rho}, in all cases, the DCE increases monotonically with respect to $n$. This implies that configurational instability grows as the excitation level increases, which explains a relatively higher frequency of experimentally detected resonances with lower radial quantum numbers. On the other hand, heavier resonances with higher radial quantum numbers have greater configurational instability, which would explain the difficulty in detecting them. For this reason, this work analyzes only a few radial quantum numbers beyond those corresponding to the meson states cataloged in PDG \cite{ParticleDataGroup:2024cfk}, avoiding a scale where mass and instability are too high, making the resonances less likely to be detected in current experiments.

\subsubsection*{Acknowledgments} RdR~thanks to The S\~ao Paulo Research Foundation -- FAPESP
(Grants No. 2021/01089-1 and No. 2024/05676-7), and to the National Council for Scientific and Technological Development -- CNPq  (Grants No. 303742/2023-2 and No. 401567/2023-0), for partial financial support. PHOS thanks to Coordination for the Improvement of Higher Education Personnel (Coordenação de Aperfeiçoamento de Pessoal de Nível Superior, CAPES - Brazil) - Finance Code 001.
The authors thank Prof. Gayane Karapetyan and Prof. Vitor de Souza for fruitful discussions. 
\bibliography{bibliography}

\end{document}